\documentclass[sigconf, nonacm]{acmart}
\AtBeginDocument{%
  }

\copyrightyear{2026}
\acmYear{2026}
\setcopyright{cc}
\setcctype{by-nc-nd}
\begin{document}

\title[Effects of Intermediate Feedback from Agentic LLM In-Car Assistants During Multi-Step Processing]{What Are You Doing? Effects of Intermediate Feedback from Agentic LLM In-Car Assistants During Multi-Step Processing}
% OG: "What Are You Doing?": Effects of Intermediate Feedback from Multi-Step Agentic LLM In-Car Assistants
% "What Are You Doing?": Effects of Intermediate Feedback During Multi-Step Processing by Agentic LLM In-Car Assistants
% "What Are You Doing?": Effects of Intermediate Feedback from Multi-Step Agentic LLM In-Car Assistants
% "What Are You Doing?": Effects of Intermediate Feedback from Agentic LLM In-Car Assistants During Multi-Step Processing

\author{Johannes Kirmayr}
\affiliation{%
  \institution{BMW Group Research and Technology}
  \city{Munich}
  \country{Germany}
}
\affiliation{%
  \institution{Augsburg University}
  \city{Augsburg}
  \country{Germany}
}
\email{johannes1.kirmayr@uni-a.de}
\author{Raphael Wennmacher}
\affiliation{%
  \institution{LMU Munich}
  \city{Munich}
  \country{Germany}
}
\author{Khanh Huynh}
\affiliation{%
  \institution{BMW Group Research and Technology}
  \city{Munich}
  \country{Germany}
}
\affiliation{%
  \institution{LMU Munich}
  \city{Munich}
  \country{Germany}
}
\author{Lukas Stappen}
\affiliation{%
  \institution{BMW Group Research and Technology}
  \city{Munich}
  \country{Germany}
}
\author{Elisabeth André}
\affiliation{%
  \institution{Augsburg University}
  \city{Augsburg}
  \country{Germany}
}
\author{Florian Alt}
\affiliation{%
  \institution{LMU Munich}
  \city{Munich}
  \country{Germany}
}

\renewcommand{\shortauthors}{Kirmayr et al.}


\begin{abstract}
Agentic AI assistants that autonomously perform multi-step tasks raise open questions for user experience: how should such systems communicate progress and reasoning during extended operations, especially in attention-critical contexts such as driving?
We investigate feedback timing and verbosity from agentic LLM-based in-car assistants through a controlled, mixed-methods study (N=45) comparing planned steps and intermediate results feedback against silent operation with final-only response. 
Using a dual-task paradigm with an in-car voice assistant, we found that intermediate feedback significantly improved perceived speed, trust, and user experience while reducing task load - effects that held across varying task complexities and interaction contexts. 
Interviews further revealed user preferences for an adaptive approach: high initial transparency to establish trust, followed by progressively reducing verbosity as systems prove reliable, with adjustments based on task stakes and situational context.
We translate our empirical findings into design implications for feedback timing and verbosity in agentic assistants, balancing transparency and efficiency.
\end{abstract}


\begin{teaserfigure}
%\vspace{-3mm}
  \includegraphics[width=1.\textwidth]{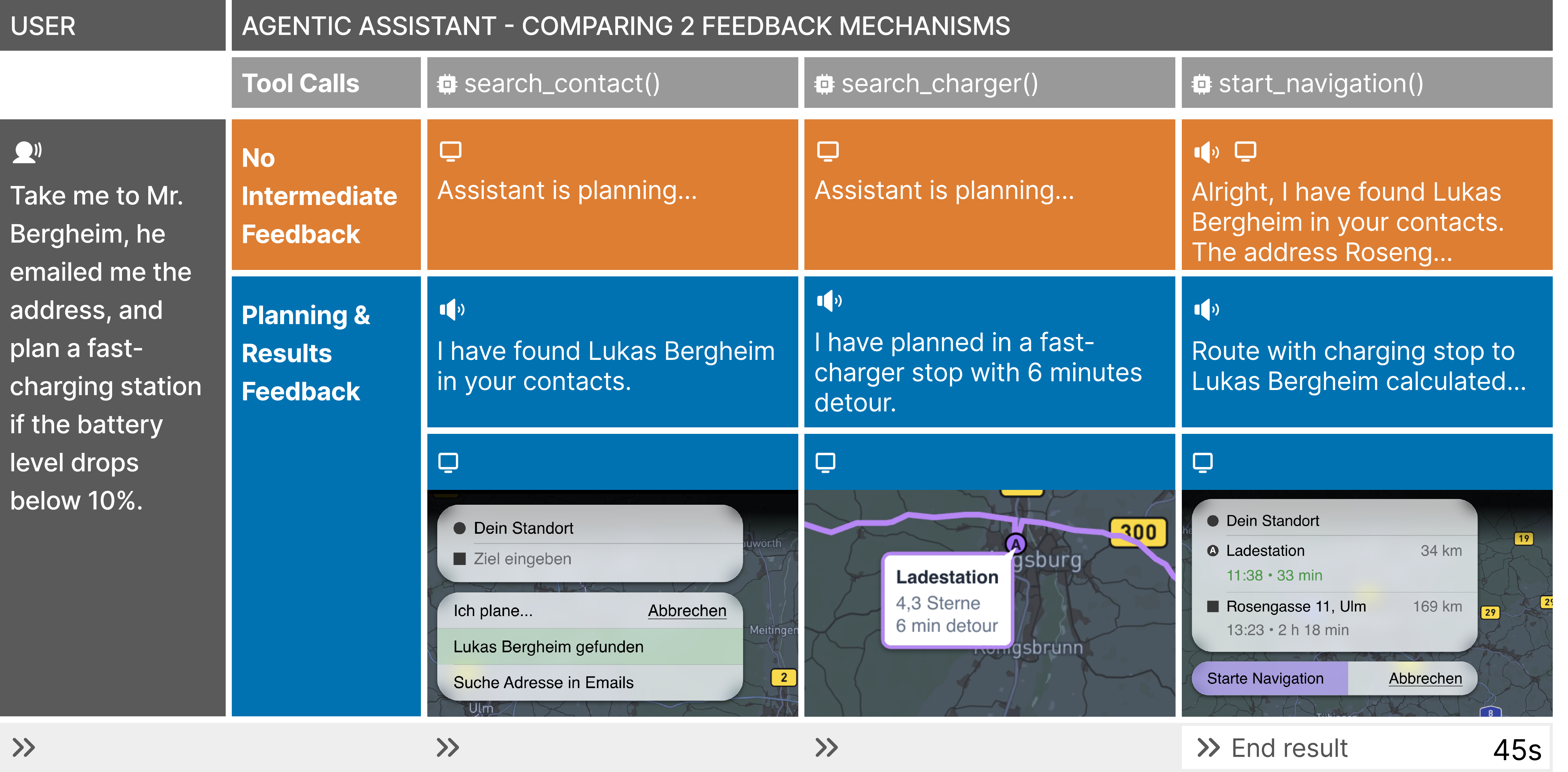}
    %\vspace{-7mm}
  \caption{How should agentic assistants communicate during long-running multi-step operations? We compare two feedback strategies: No Intermediate (NI) feedback (top), where the system acknowledges the request and remains silent until delivering the final result, versus Planning \& Results (PR) feedback (bottom), where planned steps and intermediate outcomes are communicated through synchronized audio and visual updates. The illustrated task exemplifies the multi-step operations agentic assistants perform, including contact lookup, address extraction, battery check, and charging stop planning. Our study (N=45) examines effects on perceived latency, trust, task load, and user experience in stationary and driving contexts.}
  \Description{Visualization showing two different feedback timing mechanisms studied in this work. On the left part, the multi-intent user request is displayed. From the center to the right, the assistant part is displayed. The assistant part is vertically split into three strains, the first (most upper) strain displays the tool calls executed by the assistant, the second (middle) strain shows the no intermediate feedback system, and the third (lowest) strain shows the planning and result feedback system displaying intermediate auditive and visual outputs given during assistant processing. Below the visualization is a timeline showing a total duration of 45 seconds, and indicating intermediate steps are skipped in the visualization for simplicity.}
  \label{fig:teaser}
  \vspace{5.5mm}
\end{teaserfigure}


\maketitle

\section{Introduction}
Large language model (LLM) agents are transforming how we interact with Artificial Intelligence (AI) systems, moving beyond single-turn question-answering to autonomously executing complex, multi-step tasks~\cite{acharya2025agenticai, wang_survey_2024, HOSSEINI2025100399}. 
These agentic systems decompose user requests, invoke multiple tools~\cite{schick2023toolformer}, and synthesize results across extended processing periods; from searching and comparing flight options to analyzing documents and generating comprehensive reports. 
This shift from simple query-response to autonomous task execution introduces fundamental design challenges: How should these systems communicate their progress to users during lengthy operations? When should they provide updates versus working silently in the background? How much detail is appropriate, and how should feedback adapt to different contexts and user needs?

These questions become particularly acute in dual-task scenarios where users engage with AI assistants while performing other activities \citet{stappen2023genaiautomotive}. Consider an in-car voice assistant that handles complex requests while the driver navigates traffic. 
In such context, poorly designed feedback can create dangerous distractions or cognitive overload~\cite{strayer2015assessing_cognitive_distraction_in_the_automobile, strayer_talking_2016, BEANLAND201399}, while insufficient communication can result in "ambiguous silence"~\cite{yankelovich1995ambigoussilence}, leaving users uncertain about system progress and thereby undermining trust and perceived responsiveness~\cite{porcheron2018voiceinterfaces, funkUsableAcceptableResponse2020, ueno2022trustinhumanAIinteraction}.

Currently deployed user-facing agentic systems reveal this gap through diverse feedback practices. 
Cursor~\cite{cursor}, which reached one million users within 16 months, operates nearly silently until completion with details on demand. Manus AI~\cite{manus} provides verbose step-by-step narration, while Perplexity~\cite{perplexity} previews steps but withholds intermediate results. 
This variability, from minimal to maximal transparency, highlights the lack of shared design principles for agentic feedback and underscores the need for open-research guidelines as user-deployed systems become widespread.

Decades of HCI research have established well-tested principles for designing system feedback. Yet agentic systems, with their extended processing and autonomous actions, bring new considerations.
First, latency is inherent rather than accidental.
Prior work demonstrates that unexpected delays degrade user experience~\cite{millerResponseTimeMancomputer1968, shneidermanResponseTimeDisplay1984} and that progress indicators improve perceived responsiveness~\cite{myersImportancePercentdoneProgress1985}.
However, in agentic systems, the expanded processing time is a consequence of intentionally increased reasoning and multi-step tool use, posing the question of whether findings on perceived delay mitigation transfer to waiting time as it becomes expected. 
Second, the volume of information generated during multi-step processing also expands well beyond typical single-turn interactions, raising questions about cognitive load and information distribution~\cite{bansal2024challengeshumanagentcommunication}.
As described, this is particularly critical in dual-task contexts such as driving, where even lightweight interactions can impair performance. 
Third, grounded communication requires evidence of both perception and understanding~\cite{clark191grounding, allwood1992semantics, axelsson2022modelingfeedback}. 
In long-running agentic systems, it is unclear whether grounding is maintained when perception is acknowledged upfront but evidence of understanding is deferred until the final response, or whether users expect transparency throughout the process.
Finally, trust is known to be a key factor in human–AI relationships, as it often determines whether a system is adopted and relied upon~\cite{ueno2022trustinhumanAIinteraction}. Prior work shows that transparency mechanisms such as explanations can foster trust~\cite{mehrotra2024trust, zhang2024explaining}. For agentic systems, this raises questions about which forms of feedback best support users in calibrating trust during extended and autonomous processing.

To address the gap between emerging agentic capabilities and established feedback design principles, we investigate the feedback design for an LLM-based agentic in-car assistant through a mixed-methods study (N = 45) in a controlled car simulation environment. We focus on three critical dimensions: (1) \textbf{feedback timing} — whether systems should provide intermediate informative updates during processing or deliver results only upon completion; (2) \textbf{interaction context} — how feedback strategies perform when the AI assistant is the primary versus a secondary task alongside driving; and (3) \textbf{adaptive verbosity} — how feedback detail should evolve with situational demands and long-term use.

Our quantitative experiments reveal that intermediate feedback consistently outperforms end-only delivery across multiple metrics and interaction context.
Providing planned steps and incremental results during processing increased perceived speed ($d_z=1.01$), improved user experience ($d_z=0.54$), enhanced trust ($d_z=0.38$), and, surprisingly, reduced task load ($d_z=-0.26$) despite multiple interaction points, compared to silent processing followed by a comprehensive final response.
Complementing these findings, our qualitative interview analysis uncovers sophisticated adaptation preferences. 
Participants envisioned systems that initially provide transparent, detailed feedback to establish trust, then progressively reduce verbosity in favor of efficiency as the system proves reliable.
Yet they expected transparency to immediately return for novel, ambiguous, or high-stakes requests. 
Preferences varied regarding context-aware adaptation in social settings and media consumption, with a consistent desire for simple override controls when automatic adaptation was unsatisfactory.

\paragraph{Contribution Statement.}
Our empirical findings advance agentic in-car assistant design through:
(1) \textbf{Empirical evidence} from a controlled in-car study (N=45) showing that intermediate, informative updates during extended agent operation improve responsiveness, trust, and user experience across single- and dual-task contexts;
(2) \textbf{Adaptive verbosity patterns} from qualitative interviews showing that users desire adaptive feedback detail that decreases as they experience system reliability over time, but increases situationally for novel, ambiguous, or high-stakes tasks; and
(3) \textbf{Design implications} for feedback design and adaptation in in-car single- and dual-task interaction, with potential relevance to other primary-task systems (e.g., customer support) and dual-task contexts (e.g., smart-home assistants while cooking).

\section{Related Work}\label{sec:related_work}

\subsection{Feedback Strategies in Current Agentic Systems}
Current systems vary widely from minimal (Cursor~\cite{cursor}) to verbose (Manus~\cite{manus}) feedback. 
This diversity reflects different implicit assumptions about user needs. 
Cursor's minimal approach embodies a "stay out of the way" philosophy, prioritizing uninterrupted workflow over transparency. 
This implies intermediate details may distract expert users, and suits contexts where users have established trust or where processing steps remain technical rather than decision-relevant.
Manus's verbose approach implies that transparency builds trust and helps users maintain situational awareness, though risking information overload.
Perplexity's~\cite{perplexity} hybrid strategy, showing plans but not intermediate results, attempts to balance expectations management with efficiency. 
% These divergent approaches appear driven by designer intuition rather than user research, with feedback verbosity potentially conflating domain requirements (coding vs. browsing) with user needs (expertise, trust levels).
These systems exemplify the diversity of feedback strategies deployed in practice. At the same time, they also underscore the need for empirically grounded design principles. 
With companies rarely publishing their design rationales or formative studies, the open research community lacks systematic guidance on how feedback strategies should align with varying user needs and task contexts.

\subsection{Principles from Human-AI Interaction Research}

\paragraph{Grounding and Communication.}
Research on human–AI communication builds on foundational studies of human communication and human teamwork.
Grounding communication theory~\cite{clark191grounding} highlights that effective collaboration requires maintaining common ground, meaning shared knowledge, beliefs, and assumptions. 
This common ground must be updated continuously, not only about content but also about the process of interaction.
\citet{brennan_grounding_1998} further extends this to human-computer interaction, emphasizing that people need to be able to seek
evidence that they are understood, with \citet{yankelovich1995ambigoussilence} stating that the absence of feedback leaves the user in an ambiguous silence.
In agentic systems, where a single user request initiates extended multi-step reasoning, grounding becomes more complex. The question is whether continuous intermediate updates have to be provided or if an upfront indication of perception with a deferred condensed final answer suffices to maintain common ground.

\paragraph{Latency and Waiting.}
Delays in system responses have long been shown to degrade user experience. 
Early work demonstrated that response delays decrease satisfaction~\cite{millerResponseTimeMancomputer1968, shneidermanResponseTimeDisplay1984} and that unexpected waiting increases frustration~\cite{shneidermanResponseTimeDisplay1984}. 
In voice interfaces, such delays may even lead users to assume the system has failed~\cite{porcheron2018voiceinterfaces} or that an error has occurred~\cite{funkUsableAcceptableResponse2020}. 
Nielsen highlights 10 seconds as an upper bound for keeping users' attention during waiting periods~\cite[p.135]{nielsen1994usability}.
Succesful mitigation strategies include progress indicators~\cite{myersImportancePercentdoneProgress1985}, conversation fillers~\cite{maslych2025mitigatingresponsedelays}, and explanations of ongoing processing~\cite{zhang2024explaining}.
While such strategies reduce anxiety and foster trust, their effectiveness in agentic systems, where latency is not accidental but an expected and productive aspect of multi-step reasoning, remains underexplored.

\paragraph{Trust and Transparency.}
Trust is central in human–AI interaction, influencing whether users rely on or reject system support~\cite{ueno2022trustinhumanAIinteraction}.
\citet{lee2004trust} define trust as “an
attitude that an agent will achieve an individual's goal in a situation characterized by uncertainty
and vulnerability”. In the case of agentic systems, users expose themselves to such vulnerability when delegating complex requests to the system.
For trust to emerge and persist, users must clearly understand what they can expect. Transparency is widely recognized as a means of fostering trust~\cite{mehrotra2024trust}, with explanations that align user expectations and system behavior~\cite{zhang2024explaining, liu2021inaiwetrust}. Empirical findings from human–AI collaboration confirm that transparency enhances trust, especially when systems make their reasoning explicit~\cite{vossing_designing_2022}.
\citet{hoff2015trust} further distinguishes trust in automation into dispositional trust as the person’s general tendency to trust automation, situational trust in a given context, and learned trust by prior experience. For agentic systems, situational and learned trust are particularly important, as they indicate the need for a dynamic feedback design. Importantly, in dual-task contexts such as driving, transparency must be carefully balanced: feedback must establish trust without overloading cognitive resources.

\paragraph{Human Oversight.} 
Oversight on AI systems encompasses two concepts: passive oversight (monitoring) and active oversight (human intervention) \cite{langer_effective_2024, sterz2024onthequest, almada2019humanintervention}.
Its primary goal is to enable humans to detect and correct errors in else autonomous AI decisions, which remain prone to inconsistency and limited self-awareness under real-world uncertainty \cite{kirmayr2026carbenchevaluatingconsistencylimitawareness}.
This human-AI collaboration can lead to improved agent performance \cite{he2025planthenexecute}, and user control can increase trust in the system \cite{dietvorst2018overcoming}.
However, research also highlights critical limitations: humans often overtrust plausible AI outputs or override accurate ones, undermining effective human oversight \cite{he2025planthenexecute}.
Additionally, human involvement substantially increases cognitive load on them \cite{he2025planthenexecute}.
Therefore, effective oversight requires careful design. \citet{sterz2024onthequest} emphasize that humans need epistemic access, including a sufficient understanding of what the system is doing and why.
This requires delivering the right transparency at the right time and format to support monitoring without overloading cognition, a challenge that intensifies as inter-agent communication introduces novel threat vectors in safety-critical domains \cite{stappen2026agent2agentthreatssafetycriticalllm}. This motivates our research question on designing agentic feedback for clear comprehension.

\subsection{Feedback Design under Cognitive Constraints}

\paragraph{Multimodal Feedback and Cognitive Resources}
Work on multi-modal feedback highlights strategies for distributing information across sensory channels. 
Wickens' Multiple Resource Theory suggests that tasks utilizing different perceptual modalities draw from separate cognitive resource pools, reducing interference~\cite{wickens2008multiple}. 
This principle proves particularly relevant for in-vehicle interaction, where auditory-vocal tasks interfere less with vehicle control than visual-manual ones~\cite{horrey2006examining}. However, modality selection involves trade-offs: auditory feedback preserves visual attention but lacks persistence~\cite{brewster1994guidelines}, while visual feedback provides spatial detail but requires potentially dangerous gaze shifts~\cite{burnett2013visual}. \citet{oviatt2000multimodal} shows that coordinating modalities, using audio cues to direct attention to visual information, can reduce cognitive load compared to single-modality approaches. 
For agentic systems with extended multi-step feedback, distributing information across modalities over time without overwhelming any single channel becomes a central design challenge.

\paragraph{Secondary Task Distraction in Driving and Automotive Interface Design}

Even lightweight activities can impair driving performance: hands-free conversations shrink the functional field of view~\cite{atchley2004conversationlimitsfieldofview}, and voice-based interactions impose moderate cognitive workload~\cite{strayer2015assessing_cognitive_distraction_in_the_automobile, strayer_talking_2016, strayerAssessingVisualCognitive2019}. 
Driver distraction is a major safety factor in crashes~\cite{BEANLAND201399}, yet drivers consistently underestimate its impact~\cite{white_risk_2004}.
Consequently, automotive HCI research has focused on designing assistants that minimize distraction through careful modality choice.
Speech interfaces have been found to generally outperform visual-manual~\cite{lo2013developmentofspeechinterfaces} and touch- or gesture-based~\cite{ZHANG2023103958} alternatives on driving performance measures.
\citet{braun_visualizing_2019} show that still visualizing natural language alongside speech improves information recall through text summaries, while keywords reduce cognitive load and icons increase attractiveness. 
Nevertheless, even basic voice interactions, such as dictating text messages, elevate cognitive workload compared to undistracted driving~\cite{LOEW2023106898}.
More recently, researchers have begun investigating LLM-powered conversational agents in vehicles.
\citet{sorokin2025collaborating} examine multimodal systems combining voice and graphical interfaces, arguing that maintaining mutual understanding between the user and LLM requires bidirectional translation: informing users of the AI's focus and changes while making user actions legible to the model.
The importance of shared visual context in such settings is further underlined by work on spatial referencing in in-car voice interfaces \citep{khanh2025spatial}.
However, existing studies largely address short, reactive interactions: command execution, clarification dialogues, or navigation updates. 
Comparatively little empirical work examines feedback design for \emph{long-running, agentic} in-vehicle assistants that autonomously perform multi-step tasks.

Following Norman’s principle that design must align with user needs and cognitive limits~\cite{norman2002design}, agentic systems must carefully balance trust-building transparency with cognitive safety. Feedback strategies should therefore support trust development while minimizing distraction, ideally increasing responsiveness perception and overall user experience.

\section{Research Questions}\label{subsec:study:research_questions}

% Interactive LLM-based agentic systems perform long-running tasks where multiple information has to be communicated and users must wait for results. 
% Designing how and when these systems provide feedback is a core challenge: too little feedback during processing might leave users uncertain about progress and erode trust, too much feedback can interrupt ongoing activities or increase cognitive load.

% Real-world scenarios further complicate this: users might interact with the system as a primary interaction or as a secondary interaction (e.g., while driving a car or cooking in the kitchen). 
% Additionally, processing varies in duration corresponding to task complexity, and user expectations are likely to evolve over time.
To mitigate the latency of multi-step processing and the larger volume of information within a single assistant turn, feedback must be designed to balance responsiveness, trust, grounded communication, and task load.
% This challenge becomes especially acute in dual-task contexts such as driving, where added cognitive load from poorly designed communication can impair attention and performance.
% To guide the design of feedback mechanisms, we address three research questions within an in-car assistant environment:
% \begin{itemize}
%     \item[RQ1] Feedback Timing: How does feedback timing of agentic systems (intermediate with planned steps and results vs. end-only after full processing) influence perceived latency, user experience, task load; and how should feedback timing be adapted over time and in situational context?
%     \item[RQ2] Task Duration and Interaction Context: How does processing duration and interaction context (single-task: standing vs. dual-task: driving) shape user perceptions of feedback timing and verbosity effects?
%     \item[RQ3] Feedback Verbosity: How should feedback verbosity adapt over time (longitudinal adaptation) and across situational contexts (contextual adaptation) to balance efficiency, transparency, and user trust?
% \end{itemize}
To guide the design of feedback mechanisms, we address three research questions within an in-car assistant environment:
\begin{itemize}
    \item[RQ1] When should agentic systems provide feedback: how does the timing of feedback—providing updates during task execution versus only at completion—affect users' perception of waiting time, overall experience, trust, and cognitive workload; and how should feedback timing be adapted over time and in situational context?
    \item[RQ2] How do task complexity and driving demands influence feedback preferences: how do longer processing times and whether users are actively driving shape preferences for when and how much system feedback to receive?
    \item[RQ3] How detailed should system feedback be: how should the level of verbosity in system feedback adapt over time as users become familiar with the system, and how should it adjust based on situational context to maintain an optimal balance between keeping users informed, minimizing distraction, and building trust?
\end{itemize}

\section{User Study}\label{sec:mixed_methods_study}

To answer these research questions, we conducted a mixed-methods user study (N=45).
In the quantitative part, we perform a within-subject lab study focusing on feedback timing effects under varying conditions (RQ1, RQ2).
Our independent variables are (1) Feedback Timing, (2) Task Duration, and (3) Interaction Context.
Our dependent variables are perceived (1) Speed, (2) UX, (3) Task Load, and (4) Trust.
The quantitative study design and analysis are detailed in Section~\ref{subsec:quantitative_experiment}.
Qualitative interviews extend the measured feedback timing effects and focus on adaption preferences for feedback verbosity and timing over time and situational context (cf. details in Section~\ref{subsec:qualitative_study}).

\subsection{Apparatus}\label{subsec:apparatus}

\subsubsection{User Study Environment}\label{subsubsec:meth:in_car_environment}

\begin{figure*}[t]
    \centering
    \includegraphics[width=1.\linewidth]{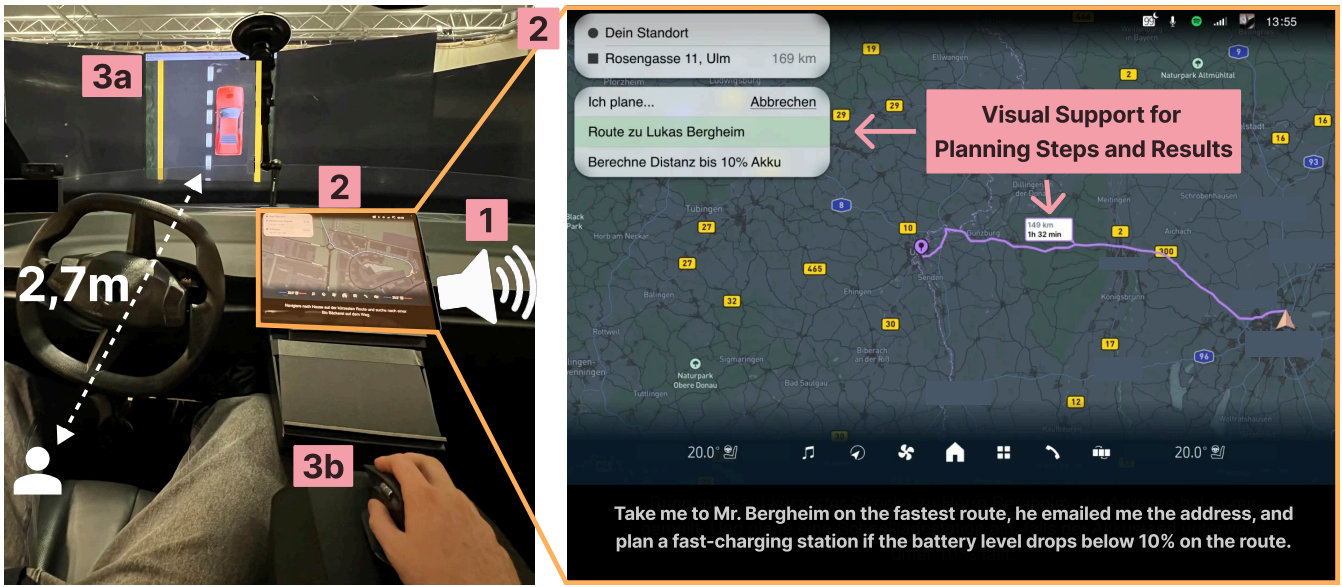}
    \caption{Study apparatus: 1. Speaker (Voice User Interface), 2. Center Display (Graphical User Interface), 3a. Driving simulation in the form of a lane-keeping task, 3b: mouse to correct continuous lateral drift for driving simulation.}
    \Description{Study apparatus for the quantitative experiments. The figure is split horizontally into two parts, with the first showing the overall car-mockup and in-car environment, and the second showing an enlargement of the visual user interface in the center display. The visual interface shows a common in-car information display in the navigation application. The assistant processing is displayed on the top left with a dynamic user interface expanding intermediate steps. The navigation app shows a route that is found for the user request.}
    \label{fig:study_setup}
\end{figure*}

The study was conducted in a controlled car simulation environment (Figure~\ref{fig:study_setup}).
The simulation environment used a fixed position, full-frame car mockup.
The participants sat in the driving seat throughout the whole study.

The interaction setup included:  
(1) a voice user interface providing auditory feedback via an external speaker, 
(2) a graphical user interface on a tablet positioned at the typical in-car center console location, and  
(3) a lane-keeping driving simulation displayed on a vertical screen outside the car-mockup and in front of the participant at 2.7 meter distance.
During the quantitative study (cf. Sec. \ref{sec:quant_study:design}), participants interacted with the voice assistant in two interaction contexts: stationary (single-task), where they sat in the car without performing the additional lane-keeping task, and driving (dual-task), where they concurrently performed the driving-related task.
The driving-related task required participants to maintain lane position by continuously correcting lateral drift.
As the steering wheel in the car mockup (visible in Figure~\ref{fig:study_setup}) was a non-functional component, we implemented mouse-based steering input for the simulation. Participants used mouse clicks to control lateral vehicle position.
This arrangement enabled a controlled manipulation of cognitive load
and provided a consistent basis for comparing feedback preferences between single-task and dual-task
situations.

\subsubsection{LLM-based Voice-Assistant System}\label{subsubsec:meth:llm_based_system}

Our research team had previously developed a fully functional agentic LLM-based in-car voice assistant capable of handling complex, multi-step tasks with real-time intermediate feedback.
This system served as the foundation for our study, providing realistic interaction flows and response timings.

Based on this, for the user study, we created a prototype in \href{https://www.protopie.io/}{ProtoPie} and deployed it on a tablet simulating the vehicle’s center display. 
This study-specific LLM-inspired prototype ensured strict comparability across all conditions in the within-subject design.
For each task configuration, the target utterance was shown on-screen for participants to read aloud. 
The system transcribed the spoken input in real-time and displayed the transcription on-screen, signaling to participants that their input had been received. 
Upon receiving the voice input, the prototype system triggered the corresponding deterministic interaction sequence, delivering visual and auditory outputs at fixed timesteps according to the configuration of one of the eight interaction tasks (cf. Figure~\ref{fig:study_design} and Figure~\ref{fig:tasks}).
Thus, unlike the working system with its dynamic LLM-generated responses, the study prototype used predefined LLM-inspired responses and fixed response timings to ensure experimental control and reproducibility, while the visible real-time transcription preserved the experience of interacting with a live system. \footnote{We provide screen videos of the \href{https://www.protopie.io/}{ProtoPie} implementation at \url{https://github.com/johanneskirmayr/agentic_llm_feedback}.}
%The interaction flows were derived from the actual system, ensuring realism, while the controlled prototype preserved naturalistic voice input alongside experimental control and reproducibility.

\subsection{Procedure}\label{sec:procedure}

The study was conducted in the car-mockup simulation environment in-person over two weeks with 45 participants ($\sim$60\,min each). 
Participants were recruited through a major automotive company's mailing lists and community channels across multiple departments to ensure demographic diversity and varying levels of familiarity with LLMs and voice assistants.
The study's procedure consists of three phases: preparation, task execution with interleaved questionnaires, and post-task interviews.

\paragraph{Preparation.}  
Participants first completed informed consent and a demographic questionnaire covering age, gender, and familiarity with: LLMs, general voice assistant systems, and the company's in-car voice assistant. 
The experimenter then introduced the driving simulation and center display prototype, with participants training on the lane-keeping task using mouse clicks to maintain lane position.
Finally, participants were briefed on the capabilities of an agentic in-car assistant and informed that the study focused on \emph{feedback delivery methods}, not AI performance.

\paragraph{Task Execution.} 
Participants then completed eight tasks covering all experimental conditions. 
Questionnaires were interleaved at different points during the session to capture the dependent variables, including perceptions of speed, task load, user experience, and trust (details in Section~\ref{subsec:quantitative_experiment}).

\paragraph{Post-Task Interview.} 
Finally, participants were interviewed about feedback adaptation preferences through three open-ended questions with follow-up prompts as needed (details in Section~\ref{subsec:qualitative_study}).

\paragraph{Ethics.} While formal approval from an ethics review board was not required in the jurisdiction where the study was conducted, all procedures adhered to recognized ethical research practices and the ACM Code of Ethics, including clear participant information, informed consent, and data protection.

\subsection{Quantitative Study}\label{subsec:quantitative_experiment}

\begin{table*}[t]
\centering
\caption{Independent variables (IVs) with condition levels and descriptions. The total completion time is defined as the moment when the assistant finishes presenting its final response.}
\label{tab:ivs_detailed}
\resizebox{\linewidth}{!}{%
\begin{tabular}{llp{.7\linewidth}}
\toprule
\textbf{IV} & \textbf{Condition} & \textbf{Description} \\
\midrule
Feedback Timing & NI & No Intermediate Feedback: Only elaborate final response after task completion. \\
    & PR & Planning\&Result: Intermediate feedback for planned steps and intermediate results during processing and summarized response after task completion. \\
\midrule
Task Duration & Medium & Task with 3 assistant steps and medium total completion time (26\,s). \\
         & High   & Task with 6 assistant steps and high total completion time (45\,s). \\
\midrule
Interaction Context  & Stationary & Single-activity: user interacts with system without concurrent tasks. \\
         & Driving  & Dual-activity: user interacts with system as a secondary activity while performing a driving-related task. \\
\bottomrule
\end{tabular}%
}
\end{table*}

\begin{figure*}[t]
    \centering
    \includegraphics[width=.9\linewidth]{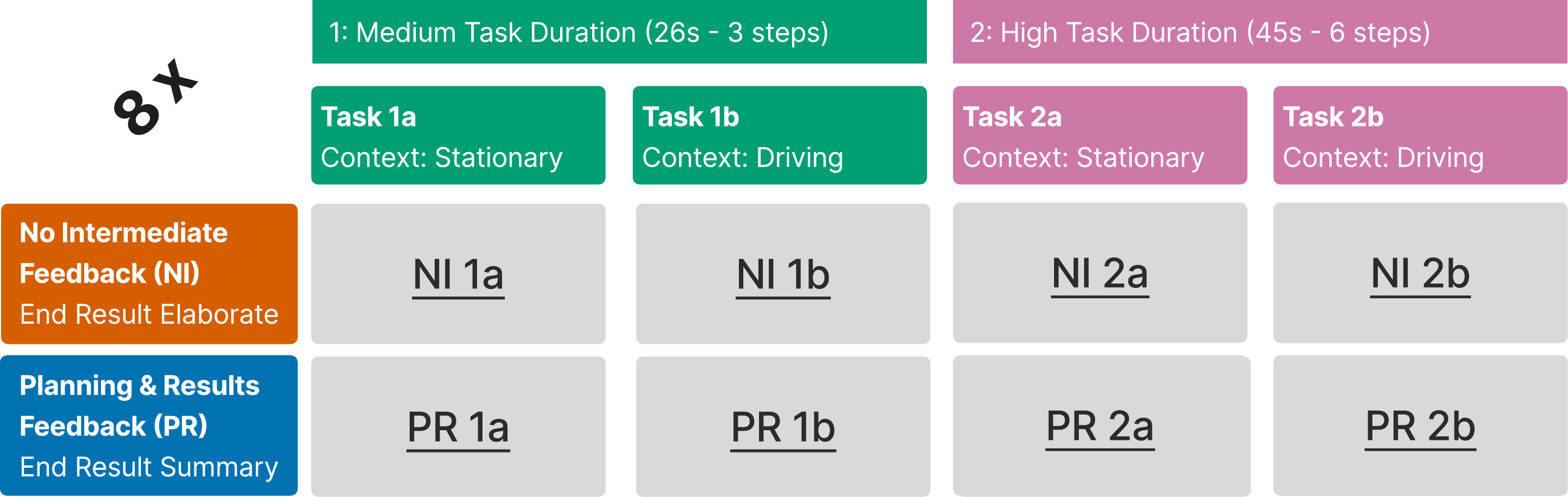}
    \caption{Quantitative study design: Each participant completed 8 tasks across the 2x2x2 conditions.}
    \Description{Visualization showing the study design with the independent variable conditions. The visualization shows 8 tasks. The task are shown in a 4 (columns) times 2 (rows) grid. The main condition of feedback timing is splitted across the rows. The condition of task duration is splitted across the first two column block (1+2) and the second two-column block (3+4). Within each two column block the context condition is splitted.}
    \label{fig:study_design}
\end{figure*}

\begin{figure*}[t]
    \centering
\includegraphics[width=1\linewidth]{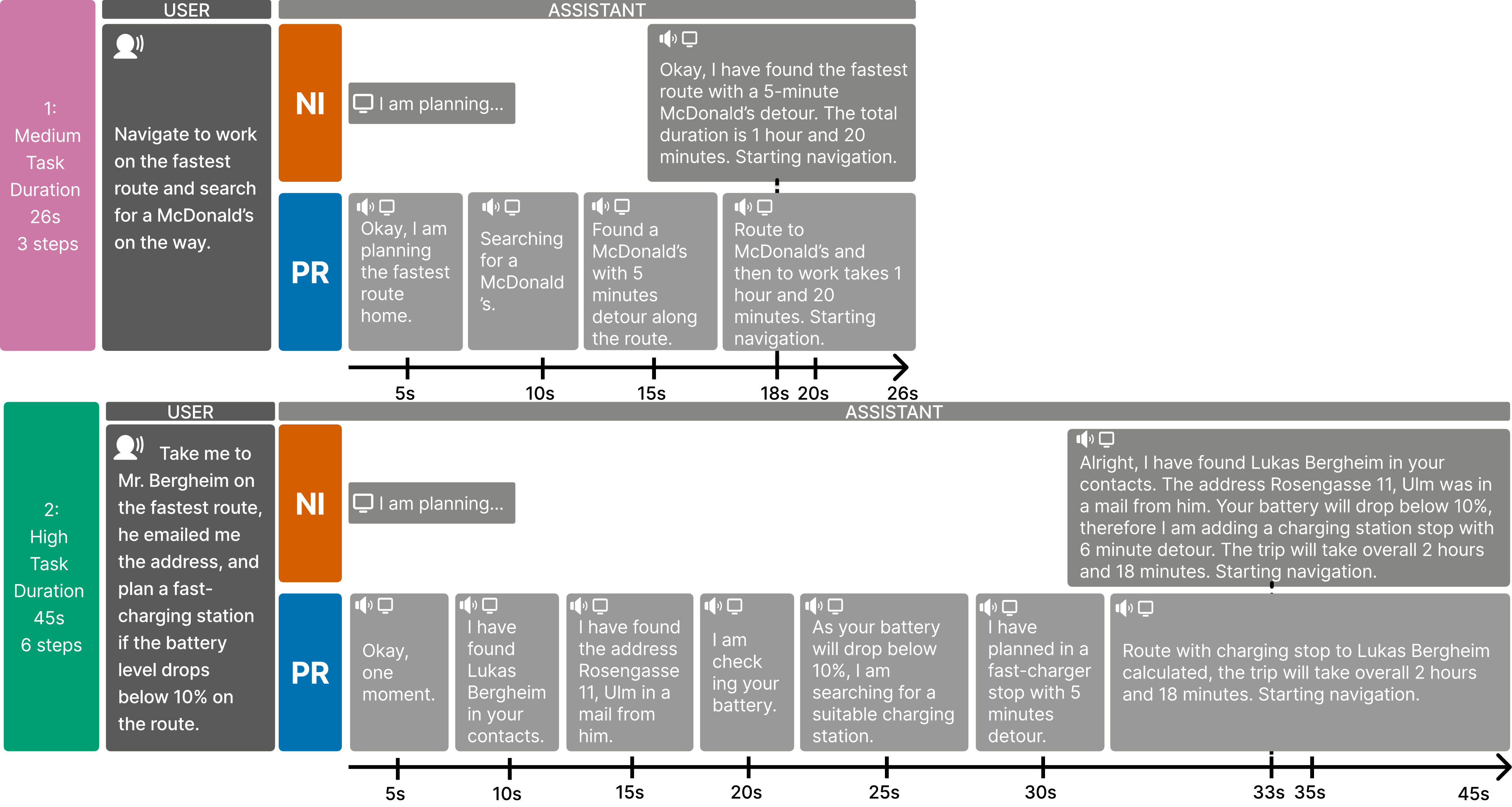}
    \caption{Quantitative study tasks: User requests for the two tasks with different durations, along with the assistant's final responses for the No Intermediate (NI) feedback system and assistant updates \& final response for the Planning \& Results (PR) feedback system. 
    As the final answer is longer for the NI feedback system, it was started earlier (at 18s compared to 20s, respectively at 33s compared to 35s) so that the last output is at the same time for both systems. 
    Note that at the beginning of both systems (NI and PR), a clicking sound, accompanied by the visual message "I am planning," is played to indicate the perception of the user request.}
    \Description{Visualization showing the user request and assistant answers for the two task durations. The visualization is split vertically into two parts. The first part shows the first task of medium duration. The second part shows the second task, which has a high duration. For each part, the user request is displayed on the left, followed by the assistant's updates and responses, which vary for the two feedback systems. The assistant's answers are divided vertically for each part to show the differences in responses for each feedback system. The upper part shows the no-intermediate feedback mechanism displaying only a visual request perception indication and a final response in the last step. The lower part shows the content of the intermediate updates for the planning and result feedback system step by step. The x-axis of the visualization features a timeline with second labels indicating at which time each intermediate update or final response is read out and visually displayed.}
    \label{fig:tasks}
\end{figure*}

\begin{table*}[t]
\centering
\caption{Dependent variables (DVs) with measurement details.}
\label{tab:design}
\begin{tabular}{llll}
\toprule
\textbf{DV} & \textbf{Measurement} & \textbf{Instrument} & \textbf{Design} \\
\midrule
Perceived Speed & After every task & 7-pt Likert & \(2 \times 2 \times 2\) \\
Overall Experience & After 2-task block & UEQ+ subset & \(2 \times 2\) (Timing × Context) \\
Task Load & After 2-task block & NASA-TLX subset& \(2 \times 2\) (Timing × Context) \\
Trust & After 4-task block & S-TIAS & Paired (Timing) \\
\bottomrule
\end{tabular}
\end{table*}

\begin{figure*}[t]
    \centering
    \includegraphics[width=.9\linewidth]{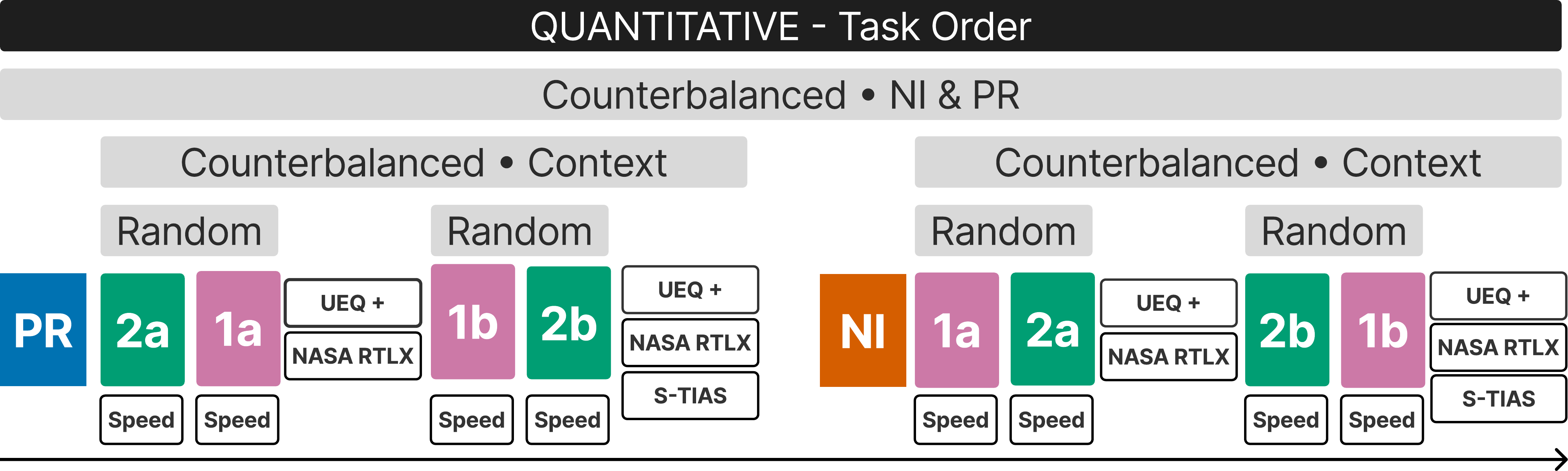}
    \caption{Task order and measurement timing: each participant performed all 8 tasks in hierarchically counterbalanced and randomized order. Dependent variables were measured (white boxes with black border) either after each task (perceived speed), after a 2-task block (UEQ+: user experience, NASA RTLX: task load), or after all 4 tasks per feedback system (S-TIAS: trust).}
    \Description{Visualization of the counterbalancing and randomization of task order. It shows the eight tasks participants perform in an exemplary order. It also shows at which time the specific questionnaires are conducted. Below the visualization, an arrow from left to right visualizes the order direction.}
    \label{fig:study_procedure}
\end{figure*}

\subsubsection{Quantitative Experiment Design}\label{sec:quant_study:design}

To capture the influence of feedback timing, task length and attentional context, we employed a controlled within-subjects \(2 \times 2 \times 2\) factorial design. 
The independent variables (IVs) and condition levels are presented in Table~\ref{tab:ivs_detailed} and further explained in the following paragraphs.

As a result, shown in Figure~\ref{fig:study_design}, each participant performs 8 tasks across the \(2 \times 2 \times 2\) conditions. 

%For the different task durations there were given two user requests:

\paragraph{IV1: Feedback Timing.} To study the effects of feedback timing, we contrasted two complementary system behaviors. 
The \textbf{No Intermediate (NI)} feedback condition served as a strong baseline: after the participant’s request was spoken, the system confirmed perception with a clicking sound and a visual message “I am planning…” on the center screen, but then remained silent until delivering the final response. 
This setup reflects how many AI agents operate and is particularly relevant in dual-task settings such as driving, where silence and background operation reduce interruptions and let the assistant “get out of the way”.
In contrast, the \textbf{Planning \& Results (PR)} feedback condition represented a transparent approach, providing informative intermediate updates during processing in addition to a final summary. 
This condition aims to mitigate uncertainty during waiting and to distribute the increased information load across smaller, progressive steps rather than presenting it all condensed at once.
Intermediate updates are conveyed auditorily and complemented by a visual display, which has been shown to be an effective design choice~\cite{braun_visualizing_2019}.
We also considered including a third condition with minimal progress cues (e.g., simple signals or conversation fillers, such as "working on it"). 
Prior work has already repeatedly shown that such cues improve perceived latency and user experience~\cite{maslych2025mitigatingresponsedelays, myersImportancePercentdoneProgress1985}. 
Given the already large design space and the novel dimension of increased information load to be conveyed, we chose not to include this condition here; instead, we revisit whether minimal cues could suffice in the discussion (Section~\ref{subsubsec:disc:minimal_cues}).

\paragraph{IV2: Task Duration.}
As a second independent variable, we varied task duration by designing two tasks with different complexity: a \textbf{medium}-duration task with 3 intermediate steps and a \textbf{long}-duration task with 6 intermediate steps.
Figure~\ref{fig:tasks} illustrates the two tasks along with the content and timing of intermediate updates and final responses for the respective feedback timing conditions.

As displayed, intermediate updates for PR were presented at fixed 5-second intervals; this corresponded to the empirical average step duration in our agentic in-car assistant prototype across different LLM models (GPT-4o~\cite{openai2024gpt4ocard}, Claude-Sonnet-4-Thinking~\cite{sonnet_system_card}, Gemini-2.5-Flash-Thinking~\cite{comanici2025gemini25pushingfrontier}). 
This interval is also below the 10-second threshold identified by~\cite{nielsen1994usability} as the upper limit for maintaining user attention.
We provide screen videos of the \href{https://www.protopie.io/}{ProtoPie} implementation of the visual and auditive feedback for the different tasks in the supplementary material. 
To avoid confounding effects of verbosity in the timing study, the appropriate level of detail for feedback in the two task examples was determined in a pre-study with (N=7) institutional HCI experts working on user experience in the LLM-based in-car voice assistant.
Across the eight tasks, interchangeable attributes (e.g., [fastest | shortest] route, [McDonald’s | Bakery], [20\% | 10\%] battery) were permuted to minimize task memorization and repeated answers, while keeping the tasks conceptually equivalent.
The user requests given to participants were deliberately explicit, hinting at the number and type of steps the assistant would take. 
% Explicit requests therefore helped align mental models and supported comparability across participants and feedback conditions independent of technical familiarity.
Explicit requests aimed to reduce variability in expectations across participants with different technical backgrounds.

\paragraph{IV3: Interaction Context.}
As a third independent variable, we varied the interaction context between a single-task and a dual-task setting.
In the \textbf{stationary} (single-task) condition, participants interacted with the voice assistant as their sole task while sitting in the stationary car-mockup.
In the \textbf{driving} (dual-task) condition, participants additionally performed the driving-related lane-keeping task described in Section \ref{subsubsec:meth:in_car_environment}, while the voice-assistant interaction remained unchanged.
The vehicle did not physically move; instead, the task simulated core attentional demands of driving.
In this setting, the lane-keeping task naturally takes priority, relegating the voice assistant interaction to a secondary task.
We included this manipulation as the concurrent task is expected to reduce available cognitive resources for processing feedback and may make waiting periods feel less idle, both potentially influencing preferred feedback timing and perceived system speed.
We thus conceptualize this manipulation primarily as a single- versus dual-task comparison rather than an investigation of vehicle motion. 
While an alternative design, comparing manual with automated driving, could isolate task demands from perceived motion, our conditions capture ecologically valid scenarios: pre-trip navigation setup versus in-drive query.

\subsubsection{Dependent Variables and Measurement}\label{sec:dv_measurements}

To capture the effects of feedback timing, task duration, and context, we measured four dependent variables (DVs) that reflect responsiveness, workload, user experience, and trust:

\begin{itemize}
    \item \textbf{Perceived Speed (DV1):} A single custom 7-point Likert item (1 = very slow, 7 = very fast) (cf.~\cite{zhang2024explaining}). This directly measures perceived responsiveness, the most immediate effect of feedback timing and task duration.
    \item \textbf{Task Load (DV2):} Three NASA-TLX~\cite{HART1988139} subscales (\textit{Mental Demand}, \textit{Temporal Demand}, \textit{Frustration}) on a 0–100 scale (0 = very low, 100 = very high). These capture cognitive demands and frustration, particularly relevant in dual-task settings such as driving. Physical Demand, Performance, and Effort were excluded because tasks were non-physical and largely reactive. It should be noted that we report the unweighted average of the included subscales (Raw TLX, RTLX~\cite{hart2006nasa-tlx-20-years-later}); this allows for comparisons across our experimental conditions, due to subscale exclusion it should not be compared directly to full NASA-TLX scores in other experiments.
    \item \textbf{User Experience (DV3):} Three UEQ+~\cite{ueqplus} subscales (\textit{Attractiveness} (overall impression), \textit{Dependability} (perceived control and predictability), and \textit{Risk Handling} (ability to detect and handle risks)) on a 7-point scale ($-3 =$ strongly negative, $+3 =$ strongly positive).
    These capture pragmatic and hedonic qualities of real-time feedback, allowing us to assess overall acceptance and perceived control.
    Subscales are reported individually, and an overall UX score was computed as a KPI, weighted by participants’ self-reported importance following UEQ+ guidelines.
    \item \textbf{Trust (DV4):} The short form of the TIAS, S-TIAS~\cite{mcgrathMeasuringTrustArtificial2025}, which covers \textit{Confidence}, \textit{Reliability}, and \textit{Trustworthiness} on a 7-point scale (1 = not at all, 7 = extremely) and focuses on trust in artificial intelligence. Trust was measured to capture how different feedback strategies influence users’ confidence in the assistant’s behavior.
\end{itemize}

Table~\ref{tab:design} summarizes the DVs, including when they were measured and the factorial structure applied in analysis.

By design, \emph{User Experience} and \emph{Task Load} were collapsed across the condition of task duration.
Although task length may influence these measures, our focus was on comparing feedback systems across contexts; duration was included primarily to span a plausible range of complexity and duration, with detailed effects analyzed only for perceived speed, as this is the most sensitive measure of waiting.
Similarly, \emph{Trust} was measured once per feedback timing condition as we targeted system-level trust (confidence/reliability/trust in assistant) rather than context-specific preferences. 
Testing differences between the study interaction context condition would provide limited generalizability given numerous possible contexts (stressful driving, media consumption, etc.) beyond our study scope. 
Notably, participants experienced both interaction contexts before providing their trust judgment, allowing them to form holistic system-level trust assessments. 
For the specific context variance included in our study, we expected user experience measures to provide more insightful metrics for context-related preferences.
This design also reduces questionnaire burden.

\subsubsection{Task Order and Measurement Timing}\label{sec:task_order}

Participants completed all eight experimental conditions in a hierarchically counterbalanced order: (1) by feedback system (NI vs. PR), (2) by context (stationary vs. driving) within each system, and randomized (3) by task duration (medium vs. high) within each 2-task block. 
As shown in Figure~\ref{fig:study_procedure}, perceived speed was rated after each task, task load and user experience were measured after each two-task block, and trust was measured once per feedback system after four tasks.

\subsubsection{Quantitative Analysis}\label{sec:quantitative_analaysis}

We use repeated-measures ANOVAs with within-subject factors \textit{Feedback Timing} (NI, PR), \textit{Context} (Stationary, Driving), and, where applicable, \textit{Task Duration} (Medium, High) or \textit{Subscale}. 
Planned paired \textit{t}-tests are used to unpack main and interaction effects. 
When multiple cell-wise comparisons are tested (Perceived Speed across Duration~$\times$~Context), we apply a Holm correction.
We report \textit{F}, \textit{p}, partial eta squared ($\eta^2_p$) for ANOVAs and \textit{t}, 95\% CI, and Cohen’s $d_z$ for paired contrasts. 
Additionally, for multi-item measures for one scale, internal consistency is assessed using Cronbach's~$\alpha$.

\subsubsection{Quantitative Hypotheses}\label{subsec:study:hypotheses}

Based on prior work on latency, cognitive load, and trust in interactive systems, we formulate the following hypotheses for our quantitative study:

\begin{itemize}
    \item \textbf{H1:} Feedback Timing Effects (RQ1): PR feedback \textbf{(a)} increases perceived speed compared to NI, \textbf{(b)} increases subjective task load compared to NI, \textbf{(c)} improves user experience, \textbf{(d)} increases user trust compared to NI.
    \item \textbf{H2:} Interaction Context and Task Duration Effects (RQ1, RQ2): \textbf{(a)} Driving context increases subjective task load compared to stationary and \textbf{(b)} longer task duration decreases perceived speed.
    %\item \textbf{H3:} Interaction Effects (RQ1, RQ2): \textbf{(a)} Context moderates the effects of PR feedback and \textbf{(b)} duration moderates the effects of PR feedback.
\end{itemize}

\subsection{Qualitative Study}\label{subsec:qualitative_study}

\subsubsection{Semi-structured Interviews.}
We complemented the quantitative experiment with semi-structured interviews to explore how participants envision adaptive feedback systems over time (RQ3) and to contextualize findings on feedback timing, task complexity, and cognitive load (RQ1–RQ2). 
After completing the tasks, participants answered three open-ended questions (translated from German):
\begin{enumerate}
    \item How much verbal feedback would you like from the system? Consider the driving situation, passengers, music, and other distractions.
    \item Should the system notify you when it is uncertain, or decide autonomously? If notified, how should this be communicated?
    \item Which system behaviors or experiences would foster long-term trust?
\end{enumerate}
Follow-up prompts were used as needed to clarify or elaborate on participants’ responses.

\subsubsection{Qualitative Analysis}\label{subsubsec:meth:qualitative_analysis}
We transcribed the audio recordings of the 45 semi-structured interviews and cleaned the data. 
We then analyzed the transcripts using thematic analysis~\cite{blandford2016qualitative} and \href{https://atlasti.com/}{Atlas.ti}. 
Two researchers independently open-coded a random subset of 20\% of the data. 
During this phase, the focus was on maintaining close proximity to participants' language and experiences while generating granular codes such as \textit{interruption reduction during media interaction} or  \textit{mute on demand}.
The researchers then met in person to compare their initial codes, resolve discrepancies, and consolidate overlapping codes.
Through this discussion, a shared codebook was developed consisting of 18 codes organized into preliminary conceptual groupings. 
Using this codebook, we divided the remaining transcripts equally for coding. 
Finally, the researchers reconvened to review coded extracts, refine code groups, and iteratively develop overarching themes. 
The researchers refined conceptual boundaries, distinguishing, for example, \textit{external real-time adaptation} (media/social context) from \textit{internal real-time adaptations} (task ambiguity, task novelty).
This resulted in five themes capturing participants’ preferences and rationales for feedback timing, verbosity, and adaptive feedback\footnote{The final codebook and theme structure are included at \url{https://github.com/johanneskirmayr/agentic_llm_feedback}.}.

\begin{table*}[t]
\centering
\caption{Participant demographics and familiarity (N=45). LLM = Large Language Model, VA = Voice Assistant.}
\label{tab:participants}
\begin{tabular}{ll}
\toprule
\textbf{Category} & \textbf{Levels and Distribution} \\
\midrule
Gender & Male: 64\%, Female: 36\% \\
Age & 18–24: 16\%, 25–34: 44\%, 35–44: 22\%, 45–54: 13\%, 55–64: 4\% \\
LLM Familiarity & 1=not: 4\%, 2: 24\%, 3: 24\%, 4: 29\%, 5=extremely: 18\% \\
VA Familiarity & 1=not: 0\%, 2: 20\%, 3: 40\%, 4: 38\%, 5=extremely: 2\% \\
In-car VA Familiarity & 1=not: 36\%, 2: 29\%, 3: 20\%, 4: 11\%, 5=extremely: 4\% \\
\bottomrule
\end{tabular}
\end{table*}

\subsection{Participants}\label{subsec:res:participants}

Table~\ref{tab:participants} shows the participants' distributions.
We recruited 45 participants (29 male, 16 female) from an automotive company, all of whom were above the age of 18. Age distribution was diverse, spanning 18–64 years, and covered the typical age range for early adopters of in-car voice assistants.
Participants reported varying familiarity with LLMs, voice assistants (VA), and in-car assistants. 
While most were familiar with general VAs (80\%), familiarity with LLMs and the company voice assistant was lower, representing an expected sample regarding prior experience, given the duration of existence of general VAs and LLMs.

\subsection{Limitations}\label{subsec:meth:limitations}

Our work is subject to the following limitations.  
First, participants were recruited from a single automotive company. Although recruitment spanned multiple departments with diverse demographics and varying familiarity with LLMs and voice assistants, reflecting the intended customer base, generalization should be done cautiously.  

Second, the driving context was simulated using a standardized lane-keeping task. This provided a consistent cognitive load increase across participants but cannot fully capture the variability of real-world driving, such as dynamic traffic or environmental distractions. Moreover, this manipulation inherently conflates perceived vehicle state with task demands. We hypothesize that the observed effects are primarily driven by the attentional demands of the concurrent task rather than perceived vehicle motion, since these demands directly constrain resources available for processing voice assistant feedback; future work comparing manual with perceived automated driving could empirically disentangle these factors. Similarly, intermediate feedback was provided at fixed 5,s intervals to isolate the effect of feedback timing; adaptive or context-aware feedback policies were beyond the scope of this controlled experiment. 

Third, our study captured immediate reactions to feedback timing and verbosity under the conditions of stationary vs. driving and medium vs. long task duration. Longitudinal adaptation over time and contextual adaptation (e.g., adjusting verbosity to cognitive load or passenger presence) were assessed only via self-reports in the qualitative interviews, rather than behavioral data from extended real-world deployments.  

Finally, feedback was always provided simultaneously via voice and visual channels. We did not explore different modality combinations (e.g., intermediate feedback visually but not auditorily) or additional modalities (e.g., haptic cues for progress indication). Including these would have led to an unfeasibly large design space; we therefore focused on the most relevant independent variables to establish a controlled baseline before adding such complexity in future work.

\begin{figure*}
    \centering
    \includegraphics[width=1\linewidth]{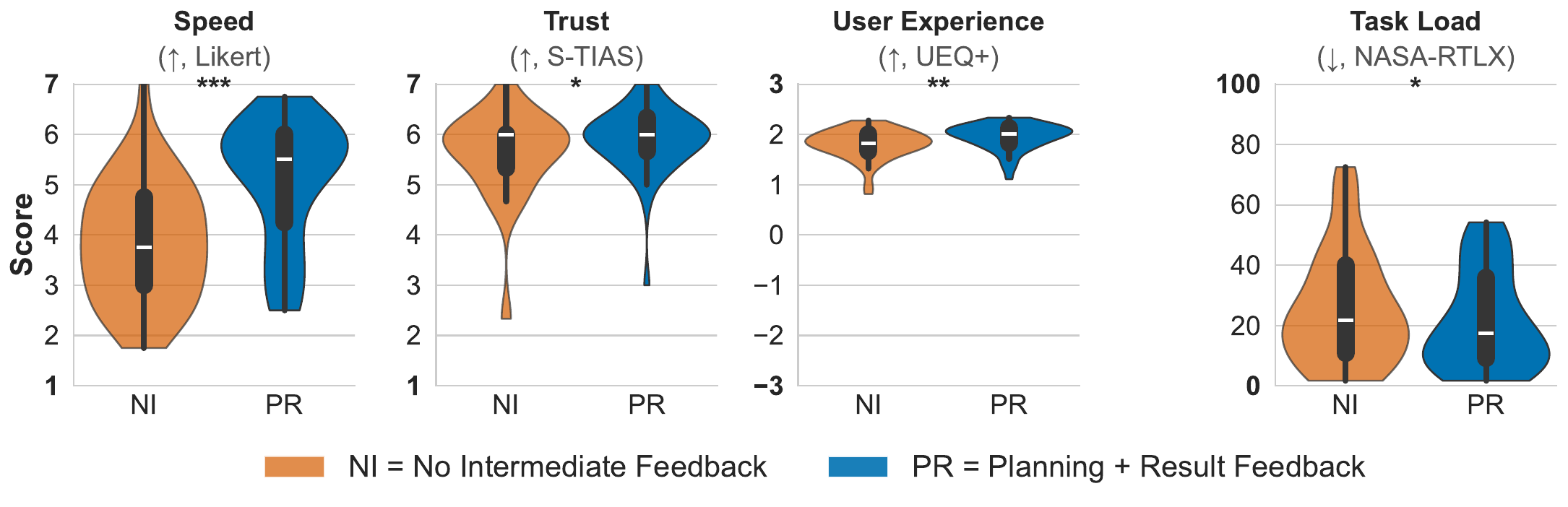}
    \caption{Scores for dependent variables of post-hoc t-tests when contrasting the feedback timing systems (NI vs. PR) collapsed across context and duration conditions. All scores show significant effect for the PR feedback timing: perceived speed shows a large effect ($p<.001, ***$, $95\%$ CI $[0.90, 1.54]$, $d_z=1.01$), user trust shows a small effect ($p=0.042, *$, $95\%$ CI $[0.01, 0.60]$, $d_z=0.38$), user experience KPI value showed a moderate effect ($p=0.002, **$, $95\%$ CI $[0.06,0.24]$, $d_z=0.54$), and task load showed a small effect ($p=0.034, *$, $95\%$ CI $[-8.54, -0.35]$, $d_z=-0.26$).}
    \Description{Scores of the dependent variables. The figure consists of four subplots arranged in a single row. All suplots show the condensed quantitative results for feedback timing (No Intermediate Feedback (NI) vs. Planning and Result (PR) Feedback) with a violin plot. The left violin plot always corresponds to the NI feedback system, the right one to the PR. The first subplot displays the perceived speed, with the y-axis representing the perceived speed score, ranging from 1 to 7. The second subplot displays the perceived trust, with the y-axis representing the trust score, ranging from 1 to 7. The third subplot displays the user experience, with the y-axis representing the user experience KPI score, ranging from -3 to 3. The fourth subplot displays the task load, with the y-axis representing the task load score, ranging from 0 to 100.}
    \label{fig:quantitative_results}
\end{figure*}

\section{Results}\label{sec:results}

\subsection{Quantitative Results}\label{subsec:res:quantitative_results}

We first present an overview of feedback timing effects across all dependent variables (DVs) in Figure~\ref{fig:quantitative_results}, followed by detailed analyses for each DV using repeated-measures ANOVAs and planned contrasts with t-tests.
The moderating effects of the interaction context and task duration conditions are presented subsequently.

Figure~\ref{fig:quantitative_results} shows a significant positive effect for the PR feedback system across all measured scores; we further unpack this in the subsequent sections.

\subsubsection{Perceived Speed (H1a, H2b)}\label{subsubsec:results:perceived_speed}

\begin{figure*}[t]
    \centering
    \includegraphics[width=.7\linewidth]{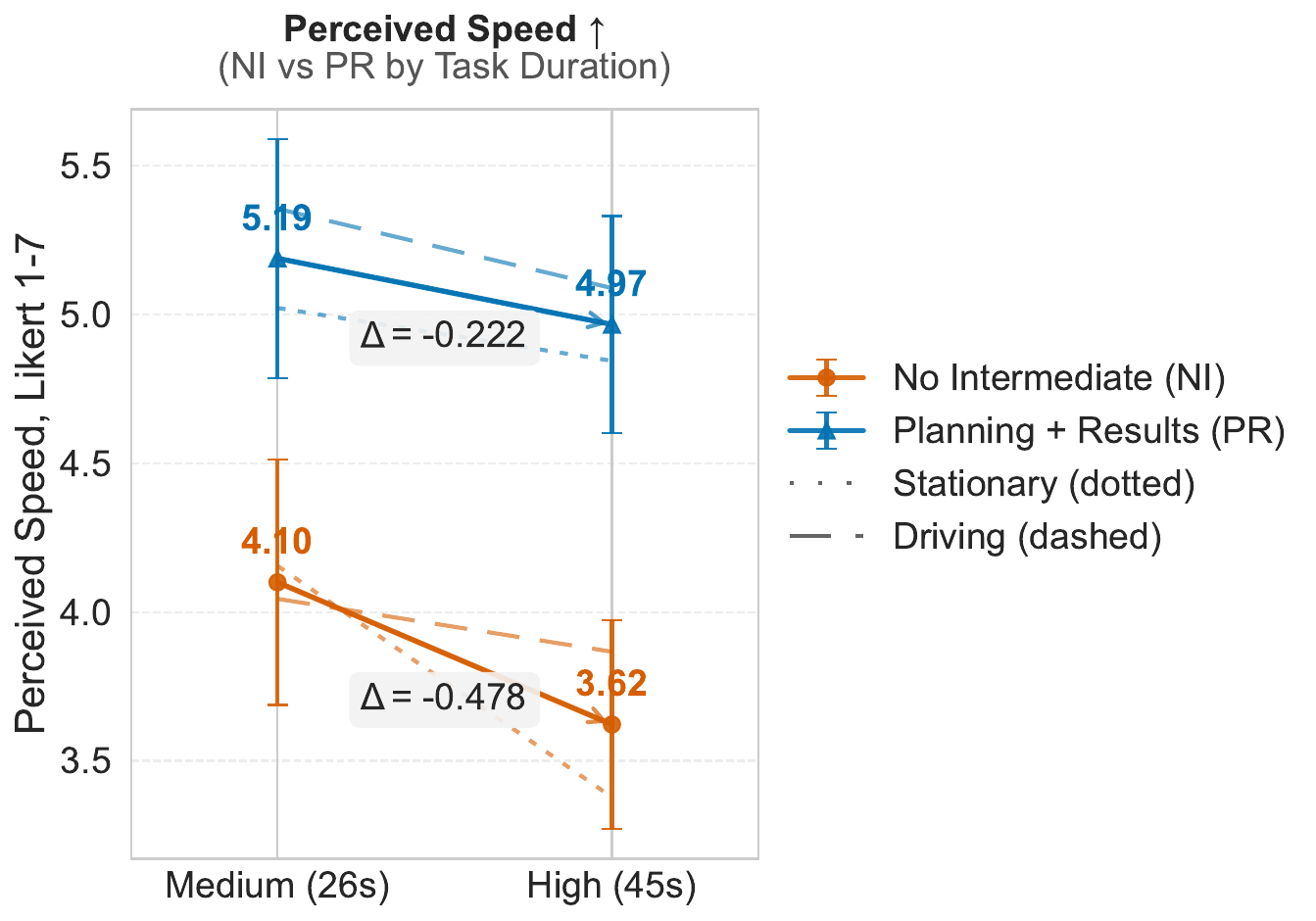}
    \caption{Perceived speed by task duration (medium, 26s vs. high, 45s) and interaction context (stationary vs. driving) for NI vs.\ PR feedback timing. Intermediate feedback reduced the negative slope in perceived speed for longer tasks.}
    \Description{Line plot showing the perceived speed results in more detail. The x-axis is split into the two task durations: medium and high, and the y-axis shows the perceived speed score. The y-axis limits the general 1 to 7 perceived speed score from 3 to 6 to enlarge the relevant lines. In the plot, the scores for each combination of feedback timing with task duration are shown as dots (4 dots). The dots corresponding to one feedback system are connected by a line along the x-axis. This shows the slope of perceived speed by the task duration along the x-axis. The plot labels that the slope for the no intermediate feedback system is $-0.478$ going from 4.10 (medium duration) to 3.62 (high duration), and that the slope for planning and result feedback system is $-0.222$ going from 5.19 (medium duration) to 4.97 (high duration). Additionally, for each feedback system, more detailed lines separating the context (stationary vs. driving) are plotted as thin overlays. This reveals that for no intermediate feedback system, the main contribution to the high negative slope comes from the stationary context.}
    \label{fig:perceived_speed}
\end{figure*}

A 2×2×2 RM-ANOVA with factors \emph{Timing} (PR vs.\ NI), \emph{Duration} (Medium vs.\ High), and \emph{Context} (Stationary vs.\ Driving) revealed a strong main effect of \emph{Timing} ($F(1,44)=58.83$, $p<.001$, $\eta^2_p=.57$).  
Collapsing across all conditions, perceived speed was significantly higher with a large effect under PR compared to NI feedback  ($t(44)=7.67$, $p<.001$, $95\%$ CI $[0.90, 1.54]$, $d_z=1.01$) (ref. Figure~\ref{fig:quantitative_results}), supporting \textbf{H1a}.
Additionally, a three-way interaction \emph{Timing × Duration × Context} emerged, $F(1,44)=4.09$, $p=.049$, $\eta^2_p=.09$.  
Planned contrasts showed that PR feedback significantly outperformed NI feedback across all Duration × Context combinations (all $p<.001$, $d_z=0.58$–$0.95$), with the largest advantage for long tasks in the stationary single-task context ($d_z=0.95$).
We also observed a main effect of \emph{Duration}, $F(1,44)=12.33$, $p=.001$, $\eta^2_p=.22$, with longer tasks reducing perceived speed, $t(44)=-3.51$, $p=.001$, $95\%$ CI $[-0.55, -0.15]$, $d_z=-0.52$, supporting \textbf{H2b}.  
Planned contrasts revealed that the reduction in perceived speed from Medium to High duration was significant under NI feedback and Stationary context ($t(44)=-3.54$, $p=.001$, $95\%$ CI $[-1.22, -0.53]$, $d_z=-0.52$).  
This implies that intermediate feedback buffered the negative impact of longer task duration on perceived speed, especially during the single-task condition (cf. Figure~\ref{fig:perceived_speed}).

\subsubsection{Task Load (H1b, H2a)}\label{subsubsec:results:task_load}
A 2×2×3 RM-ANOVA with factors \emph{Timing} (NI vs.\ PR), \emph{Context} (Stationary vs.\ Driving), and \emph{Subscale} (Mental Demand, Temporal Demand, Frustration) revealed a main effect of \emph{Timing} ($F(1,44)=4.79$, $p=.034$, $\eta^2_p=.10$), and an approaching but no significant main effect of \emph{Context} ($F(1,44)=3.96$, $p=.053$).  
Collapsing across subscales and contexts, task load was, unexpectedly, significantly lower with a small effect under PR compared to NI feedback ($t(44)=-2.19$, $p=.034$, $95\%$ CI $[-8.54, -0.35]$, $d_z=-0.26$) (ref. Figure~\ref{fig:quantitative_results}), contradicting \textbf{H1b}, which expected PR to \emph{increase} task load due to multiple interaction points.  
Further planned contrasts showed this reduction was primarily driven by the \emph{Frustration} subscale ($t(44)=-2.04$, $p=.047$, $95\%$ CI $[-12.81, -0.08]$, $d_z=-0.26$), while Mental and Temporal Demand did not differ significantly (all $p>.10$). Cronbach's $\alpha=.77$ indicated acceptable internal consistency for the composite.  
Finally, although task load tended to be higher in the driving than the stationary context, this effect did not reach significance ($p=.053$), thus providing no clear support for \textbf{H2a}.

\subsubsection{User Experience (H1c)}\label{subsubsec:results:user_experience}
A 2×2×3 RM-ANOVA with factors \emph{Timing} (NI vs.\ PR), \emph{Context} (Stationary vs.\ Driving), and \emph{Subscale} (Attractiveness, Dependability, Risk Handling) revealed a strong main effect of \emph{Timing} ($F(1,44)=12.09$, $p=.001$, $\eta^2_p=.22$), but no main effect of \emph{Context} ($p=.85$) and no \emph{Timing × Context} interaction ($p=.32$).  
Collapsing across contexts, intermediate feedback (PR) significantly improved all three user experience subscales with small to medium effect compared to final-only feedback (NI):  
Attractiveness ($t(44)=2.29$, $p=.027$, $95\%$ CI $[0.04,0.66]$, $d_z=0.38$),  
Dependability ($t(44)=2.44$, $p=.019$, $95\%$ CI $[0.07,0.72]$, $d_z=0.47$), and  
Risk Handling ($t(44)=3.90$, $p<.001$, $95\%$ CI $[0.35,1.10]$, $d_z=0.60$), with the strongest effect observed for Risk Handling.
Cronbach's $\alpha$ indicated good reliability for all multi-item measures of the subscales ($\alpha=.74$–$.86$).  
Using the UEQ+ KPI weighting by self-reported importance of each subscale, PR feedback also improved the overall user experience composite score with a medium effect size ($t(44)=3.30$, $p=.002$, $95\%$ CI $[0.06,0.24]$, $d_z=0.54$)(ref. Figure~\ref{fig:quantitative_results}), supporting \textbf{H1c}.

\subsubsection{User Trust (H1d)}\label{subsubsec:results:trust}
Trust was measured once per participant for each feedback condition after completing all tasks for that condition.
Collapsing across all contexts and durations, trust ratings were significantly higher with a small effect under intermediate feedback (PR) compared to final-only feedback (NI) ($t(44)=2.10$, $p=.042$, $95\%$ CI $[0.01, 0.60]$, $d_z=0.38$), supporting \textbf{H1d}.
Internal consistency for the three S-TIAS items (confidence in the system, reliability, and trustworthiness) was good (Cronbach’s $\alpha = 0.84$).
At the subscale level, PR feedback significantly improved Reliability ($t(44)=2.20$, $p=.033$, 95\% CI $[0.03, 0.60]$, $d_z=0.41$) and Trustworthiness ($t(44)=2.41$, $p=.020$, $95\%$ CI $[0.05, 0.61]$, $d_z=0.34$), but not Confidence ($p=.194$).
Counterbalancing ensured that overall trust scores were not affected by order.
Nevertheless, as trust can be learned through experience~\cite{hoff2015trust}, we tested for order effects.
A 2x2 mixed ANOVA with \emph{Timing} (NI vs.\ PR) as a within-subject factor and \emph{Order} (NI-first vs. PR-first) revealed no significant main effect of \emph{Order} ($p=.418$) and no significant \emph{Timing × Order} interaction ($p=.083$).

\subsubsection{Summary of Duration and Context Effects}
Perceived speed showed a significant \emph{Timing × Duration} interaction ($p=.049$), where intermediate feedback buffered the negative effect of longer tasks. 
For the other DVs, the effect of duration was not separately measured.
Interaction \emph{Context} (stationary vs.\ driving) produced no consistent main or interaction effects; a trend toward higher task load while driving ($p=.053$) did not reach significance, and no \emph{Timing × Context} interactions were observed.
Overall, intermediate feedback consistently improved perceived speed, user experience, and trust while unexpectedly reducing task load; duration mainly affected perceived speed, whereas interaction context showed no consistent effects across dependent variables.

\subsubsection{Summary of Demographic and Familiarity Effects}
No significant main effects of age or familiarity with LLMs, voice assistants, or the company's voice assistant were observed on any dependent variable.
However, moderation analyses revealed that higher LLM familiarity significantly amplified improvements in trust ($p = .017$, $d_z = 0.74$) and user experience ($p = .021$, $d_z = 0.72$) from NI to PR feedback conditions.
This pattern may reflect PR feedback's structural similarity to Chain-of-Thought \cite{wei2022cot} reasoning outputs typical of LLMs, though alternative explanations warrant investigation.
No additional moderation effects reached significance. Future research with a larger sample size may better capture the influence of demographic variables and technical familiarity.

\subsection{Qualitative Findings}\label{subsec:res:qualitative_findings}

Complementing the quantitative results, the interviews after completing all 8 tasks reveal deeper insights into why participants preferred intermediate feedback and how they envision future adaptive systems (RQ3).
Through inductive thematic analysis of 45 semi-structured interviews, we identified five major themes that contextualize our experimental results.
Below, we detail each theme, illustrating findings with representative participant quotes (P1–P45).
All quotes are translated from German into English.

\subsubsection{T1: Longitudinal adaptation should be gated by trust and enabled by learning.}
Across interviews, participants emphasized that feedback verbosity should decrease over time when the system has proven itself and learned their routines. 
Most described trust as the primary precondition for reducing transparency; some explicitly tied this to repeated successful outcomes and predictable behavior. 
As P1 noted, “\emph{Trust only grows over time-when it delivers good results consistently}”. 
Echoing this, P28 suggested, “\emph{With more trust, it can say less}”.
Participants also articulated how such reductions would be operationalized: the system should recognize recurrent tasks and remember prior choices to streamline future interactions. 
P9 commented, “\emph{For recurring complex tasks, it could keep it shorter}”. 
Similarly, P35 proposed, “\emph{It should remember my answers and ask fewer questions next time}”. 
% This suggests a trajectory from initially transparent, stepwise feedback toward more concise, efficiency-oriented interactions as user-specific patterns are learned. 
While participants generally endorsed this progression, they also implied that reduced transparency should remain reversible if trust is challenged.

\subsubsection{T2: Real-time external adaptation: balancing responsiveness with social and media context}
Participants advocated for feedback that adapts to the surrounding situation, particularly media and social contexts, yet views diverged on how much to reduce interruptions. 
Several participants preferred minimal intermediate speech while listening to music or podcasts; P39 said, “\emph{With music on, one condensed summary would be better}”. P2 added, “\emph{It’s annoying if things are said multiple times during a podcast}”. 
In contrast, others wanted consistent output regardless of concurrent media: “\emph{Even with music, I want the output when I ask for it}” (P31; see also P35, P42).
Social presence introduced additional sensitivity: some felt continuous intermediate speech could be intrusive with passengers-“\emph{With a passenger, intermediate steps could be more tiring}” (P13), and preferred a single end summary to avoid disrupting conversation (e.g., P22).
%Taken together, this implies that external context \emph{likely} matters, but there is no uniform policy acceptable to all participants. Indications are that when the assistant is a \emph{primary} activity, participants did not want reductions (“I asked, so tell me now”); when it is a \emph{secondary} activity alongside media or social interaction, many preferred fewer interruptions, leaning toward a one-time condensed response. 
%However, given the divergent preferences, this area appears underspecified and warrants further investigation.

\subsubsection{T3: Real-time internal adaptation: task ambiguity, stakes, and novelty heighten transparency needs}
Participants reached a clear consensus that ambiguity requires clarification independent of trust. 
“\emph{It should ask follow-up questions when something is ambiguous}” (P6; see also P7, P10).
They further differentiated between high- and low-stakes actions: for decisions with higher consequence or higher error cost, most participants wanted verification and intermediate checkpoints: “\emph{With contacts, it’s important to ask}” (P4); “\emph{With emails, a lot can go wrong}” (P13). 
In contrast, for low-stakes actions (e.g., choosing a fast-food stop), many preferred the assistant to proceed with minimal dialog: “\emph{McDonald’s, just take the faster one}” (P5).
Finally, participants highlighted novelty as a cue for more transparency: “\emph{As long as the information is new and not repeated, it’s relevant to me}” (P16). 
%Together, these reports suggest a principled rule-of-thumb: increase intermediate feedback when requests are ambiguous, high-stakes, or novel; compress feedback when they are clear, low-stakes, or familiar.

\subsubsection{T4: Active user control as a safety valve for feedback.}
Regardless of context policy, there was broad support for lightweight, user-driven controls to modulate verbosity. 
Participants repeatedly requested a way to mute or dampen spoken feedback when needed. 
P23 stated, “\emph{I want to choose how much information I get, mute is very important}”. 
P11 was even more direct: “\emph{I should be able to tell it not to talk}”. 
Participants suggested using such controls especially during media playback or when passengers are present (e.g., P44). 
%This suggests that active, user-facing controls can serve as a safety valve when real-time adaptation does not perfectly fit the situation.

\subsubsection{T5: Progressive chunking lightens cognitive load compared to end-only “dumps”}
Participants contrasted progressive, small updates with a cognitively heavier “all at once” delivery. 
P1 put it succinctly: “\emph{It’s the same information, but a complete dump is harder to absorb}”. 
Several echoed that stepwise updates made the process feel lighter and easier to follow than a dense end summary.
%Qualitative explanations illuminate why intermediate feedback improved perceived speed and did not raise task load. Several participants contrasted progressive, small updates with a cognitively heavier “all at once” delivery. As P1 put it, “\emph{It’s the same information, but a complete dump is harder to absorb.}” This aligns with our quantitative pattern (Section~\ref{subsec:res:quantitative_results}) and suggests that well-paced intermediate updates can distribute cognitive effort over time, making the overall experience feel faster and less frustrating, even when total completion time is unchanged.

\section{Discussion \& Implications}\label{sec:discussion}

\subsection{Feedback Timing (RQ1, RQ2)}\label{subsec:disc:feedback_timing}
\subsubsection{Intermediate feedback outperforms across conditions}

Our quantitative results demonstrate that intermediate updates substantially improved perceived speed with a large effect, user experience with a medium effect, and trust with a small effect while lowering perceived frustration and task load with small effects (cf. Section~\ref{subsec:res:quantitative_results}). 
This pattern extends decades of research on responsiveness. 
Early HCI studies showed that unexplained delays reduced perceived speed~\cite{maslych2025mitigatingresponsedelays}, responsiveness improved overall user experience~\cite{millerResponseTimeMancomputer1968}, and unexpected waiting increased frustration~\cite{shneidermanResponseTimeDisplay1984}. 
More recent work shows that explanations during delays were shown to increase trust~\cite{zhang2024explaining}. 
Our results extend these findings to the context of agentic LLM-based assistants, where delays are not just incidental but inherent to handling multi-intent requests and extended reasoning steps.
While participants were briefed on these mechanisms and given explicit requests that hinted at multi-step processing, latency may still have felt unexpected at times due to misaligned mental models ~\cite{johnson1986mental, steyversWhatLargeLanguage2025}.
% We confirm these effects in the context of agentic assistants, but under conditions where delays should in principle be expected: participants were explicitly briefed about the system’s agentic nature, and the explicit multi-intent requests made the need for extended processing salient. 
% However, full alignment of user and system mental models~\cite{johnson1986mental} cannot be assumed. 
% For example, \citet{steyversWhatLargeLanguage2025} show that users frequently overestimate LLM accuracy. 
% This implies that even though the overall benefits of intermediate feedback stand, they may in part be explained by its ability to mitigate imperfect user-system mental-model alignment.

Intermediate updates were particularly helpful for longer, more complex tasks: perceived speed declined under final-only feedback but was buffered by stepwise updates. 
This underscores the importance of sustaining a sense of progress as task duration increases. 
The strongest effects were observed when interacting without a secondary task, suggesting that idle waiting is especially pronounced. 
While benefits were also observed during dual-task interaction, the moderation effects were weaker, and for other dependent variables, we did not find significant interactions with context or duration. 
This highlights robustness but also underscores the need for in-the-wild validation under more varied dual-task demands.

\paragraph{Design implication:} From the user's perspective, agentic in-car assistants should provide intermediate updates during long-running, multi-step tasks if possible, particularly as task complexity grows. This guideline is most critical in early phases of adoption when overall trust is still developing. Benefits hold across both stationary and driving contexts, but may vary with situational trust or when competing attentional demands alter user priorities (see Section~\ref{subsubsec:dis:verbosity_situational_context_adaption}).

\subsubsection{Are simple progress cues enough, or must updates include content?}\label{subsubsec:disc:minimal_cues}
A natural design question is whether lightweight progress indicators (such as auditory fillers, e.g., "\textit{working on it}", or visual and haptic cues) could substitute for content-rich updates, especially once trust has been established.
Our results and related work suggest otherwise.

Early work on “ambiguous silence” showed that minimal feedback leaves users uncertain about what the system is doing~\cite{yankelovich1995ambigoussilence}, a problem documented in voice interfaces~\cite{porcheron2018voiceinterfaces} and reflected in theories of grounding: communication requires not just perception-level confirmation (“I heard you”) but also understanding-level evidence (“I understood what you meant”)\cite{allwood1992semantics, clark1996using, axelsson2022modelingfeedback}.
If intermediate steps are hidden until the final response, users are deprived of understanding-level feedback throughout the wait.
This forces them into laborious checking behaviors to re-establish common ground, a dynamic well described in conversational grounding research \cite{brennan_grounding_1998}.

Trust research supports this interpretation: ~\citet{zhang2024explaining} found that simply notifying users of a delay is insufficient, while providing justifications for what is happening increases trust.
Our workload results reinforce the point that when identical information is delivered in smaller steps, task load decreases significantly, with a small effect size, compared to receiving a single condensed final response.
Participant explanations (T5) complement this: they described stepwise updates as cognitively lighter than dense “dumps”. This is also important to enable effective human oversight, as \citet{sterz2024onthequest} outlines that epistemic access, including comprehending what the agent is doing, is important. 

An analogy to visual driver distraction guidelines is interesting: AAM standards restrict individual glances to in-car displays while driving to two seconds and cumulative glances to complete one task to 20 seconds~\cite{ADFTW2006, braun_visualizing_2019}.
A single long information dump parallels a prolonged glance, intensifying momentary distraction, whereas stepwise updates are akin to multiple brief glances - each less demanding and less disruptive to ongoing tasks.

\paragraph{Design implication:} Content-bearing intermediate updates appear more effective than progress-only cues. They help preserve grounding, maintain trust, and distribute cognitive effort more evenly across time. 
This holds when the cognitive feedback channel (e.g., auditive) is available; if not, updates might be experienced as interruptive, and user preferences can differ (see Section~\ref{subsubsec:dis:verbosity_situational_context_adaption}). 

%In cases where multiple agentic steps finish in parallel, designers may consider delaying or batching updates into digestible chunks: an open direction for future work.
One promising direction to gain efficiency while maintaining these benefits is through learned cue associations. 
As users learn mappings between simple cues and specific system actions, those cues can effectively convey the content of what the assistant is doing without verbose language. 
\citet{cho2025persistent} demonstrate this approach with direct multimodal feedback cues in repetitive tasks, reducing cognitive load while preserving informational value.

\subsection{Feedback Verbosity (RQ3, RQ2)}\label{subsec:disc:feedback_verbosity}

\subsubsection{Long-term verbosity adaption -- gated by learned trust through demonstrated reliability}

Participants (T1) reported that reductions in verbosity were only acceptable once the in-car assistant had demonstrated reliability.
Trust was described as the decisive factor, developing over time through repeated successful interactions.
This resonates with~\citet{hoff2015trust} three layers of trust: dispositional, situational, and learned.
In our qualitative findings, learned trust - understood as users' emerging confidence grounded in experienced reliability - was central for long-term verbosity adaption.
As the system repeatedly performed well, participants expressed willingness to accept less transparency in favor of greater efficiency.
This illustrates a trajectory where transparency, represented by intermediate feedback, initially outweighs efficiency due to its benefits for perceived speed, trust, and task load.
Over time, as learned trust grows, verbosity can be scaled back without eroding confidence in the system. 
Reversibility was mentioned (T1) by some participants, but is inherently part of this learned-trust-dependent adaptation: when reliability falters, transparency must reappear.

\paragraph{Design implication:} Our empirical findings suggest that feedback verbosity should start high to establish transparency, then decrease as the system demonstrates reliability.
This trajectory treats demonstrated reliability as a practical proxy for learned trust, enabling efficiency without undermining confidence.

\subsubsection{Real-time situational context adaptation - internal vs. external factors}\label{subsubsec:dis:verbosity_situational_context_adaption}

Participants described the need for real-time adjustments to feedback, shaped by both internal task factors and external situational factors.
For internal factors (T3), situational trust was crucial: when a task was novel or ambiguous, participants sought greater transparency, regardless of their prior experience.
This supports prior work showing that explanations are most valuable when outcomes are uncertain or likely to surprise the user~\cite{SREEDHARAN2021103558}.
Similarly, high-stakes requests such as contacting people or handling emails were seen as requiring verification, whereas routine or low-stakes actions could be handled more efficiently with minimal verbosity.
This highlights the need for transparency to scale with task stakes, enabling human control and potential intervention, which \citet{dietvorst2018overcoming} found to increase user trust .

By contrast, external context factors (T2), such as media consumption or social interaction in the car, produced divergent preferences.
Some participants preferred a condensed, final-only summary to avoid interruptions, while others valued consistent updates independent of distractions.
Because no uniform policy fits all, user-facing controls (e.g., mute button or voice command) become an effective and wanted (T4) practical way to resolve mismatches between system adaptation and individual preferences.
Although not explicitly mentioned by participants, we hypothesize that verbosity controls should also permit the expansion of detail on demand — for instance, through the visual unfolding of a step within an interface element, or in response to user-initiated clarification requests.

\paragraph{Design implication:} Our qualitative findings suggest that verbosity should be increased when task factors are novel, ambiguous, or high-stakes and can be reduced for routine or low-stakes requests. External context adaptation remains more contested: until robust personalization methods exist, systems might provide simple user controls for in-the-moment overrides.

\subsection{Applicability across domains}  
Our design implications could inform considerations for other user-facing agentic systems beyond in-car assistants, though direct transferability requires further investigation.
We identify two main user-agent interaction settings where our findings may offer relevant insights:  (i) when the agent constitutes the primary task, and (ii) in dual-task scenarios where primary and secondary activities rely on different cognitive channels. 
In contrast, when both tasks share the same channel (such as coding copilots, writing assistants, or listening to media while driving), intermediate updates may risk interference rather than relief. Since we only qualitatively found divergent preferences, we cannot yet offer implications in this regard.

In primary-task systems, such as social or service robots~\cite{lukasik2025fromrobotstochatbots, kraus2022includingsocial, song2022role} and customer-support agents~\cite{airportchatbot, chen2024travelagentaiassistantpersonalized}, our findings on adaptive feedback detail and keeping informative updates to preserve grounded communication appear particularly relevant. 
In dual-task contexts with different channels, examples include smart home assistants used during everyday activities, such as cooking~\cite{jaber2024cookingwithagents}, or emerging wearable agents that deliver feedback via audio or haptics alongside physical tasks. 
Here, along with feedback timing implications, our results on verbosity adaptation based on internal task factors and lightweight user control may help strike a balance between responsiveness and flexibility.

Regarding temporal scope, our implications target agentic systems with execution times ranging from multiple seconds to one minute. They likely do not extend to "deep agent" systems, such as OpenAI's Deep Research~\cite{openai_deep_research}, which operate over multiple minutes to half an hour. Such extended durations naturally push interactions into background processing, as sustained intermediate feedback across many minutes would likely overwhelm users rather than maintain engagement. This distinction raises an interesting research question for future work: identifying the temporal threshold where agentic systems should transition from maintaining user attention to background operation.

Finally, the application of our implications ultimately also depends on domain- and modality-specific constraints. 
Just as driving imposes limits on distractions, other domains will have their own boundaries.
Our insights may inform design considerations in other domains, but likely require adaptation to specific contextual constraints rather than direct transfer.

\subsection{Future Design and Tech Challenges}

Our empirical findings yield design implications for agentic in-car assistants, yet translating these into operational systems poses open challenges for further research.
Rather than proposing concrete solutions, we outline design challenges and point to directions and related work that may inspire future work.

\paragraph{Intermediate feedback timing and content.}
Current LLM agents operate through sequential tool calls, and intermediate outputs can be prompted if they are supported natively (e.g., OpenAI’s tool preambles~\cite{tool_preamples}), or can be supplied by an additional asynchronous LLM.
Open design challenges include coordinating overlapping intermediate voice outputs, and deciding which information to present in the informative updates versus deferring for follow-up clarification.

\paragraph{Long-term verbosity adaptation based on demonstrated reliability.}
Our empirical findings indicate that users prefer an adaptive system that reduces feedback verbosity, as they perceive sufficient reliability.
This relates to Google's PAIR guidebook on trust, which identifies ten levers of trust \cite{trustPairGoogle};
our findings specifically reveal a dynamic relationship between two of these levers: \textit{Reliance}, indicating how much users can depend on the system, and \textit{Transparency}, which helps users understand and predict agent responses.
While trust itself is a latent variable that cannot be directly measured \cite{xie2019robot, vereshak2021howtoevaluatetrust}, the relationship between the more objective demonstrated reliability from interaction history and verbosity preferences may offer a more tractable direction for future system design.
Prior work on online trust estimation, using Bayesian Networks to infer user trust states in robot-collaborative settings \cite{xu2015optimo} or Finite State Automata to model trust from acceptance behaviors \cite{VIRVOU2024120759}, may inform future operational approaches.
Trust-related behavioral signals from interaction history, such as acceptances, interruptions, corrections, rejections, and override rates, are commonly used and may serve as practical proxies for demonstrated reliability, potentially yielding systems better aligned with the preferences we identified.
Future work would need to determine which signals best estimate demonstrated reliability and identify appropriate verbosity adaptation levels.

\paragraph{Situational verbosity adaptation based on task novelty, stakes, and confidence.}
Novelty can be estimated from memory and task history~\cite{packer2024memgptllmsoperatingsystems, kirmayr2025carmem}, and stakes are often estimable in closed domains with limited action space.
In contrast, detecting ambiguity remains an open challenge and is an active research area \cite{kobalczyk2025activetaskdisambiguationllms}.
Current LLMs' internal confidence assessments are yet poorly calibrated \cite{zhu-etal-2023-calibration, kalai2025languagemodelshallucinate}, limiting reliable detection.

These challenges, informed by our empirical findings, offer concrete entry points for future systems research on agentic feedback.

\section{Conclusion}

In this work, we address the central challenge of how agentic assistants should communicate progress and manage information load during long-running tasks.
Current systems vary from silent background operation to detailed step-by-step updates - the right balance is especially critical in dual-task settings where cognitive load is constrained.
Through a controlled mixed-methods study of an in-car agentic assistant, we show that (1) intermediate informative updates improve trust, perceived speed, and user experience while reducing task load, and (2) feedback should adapt over time - starting transparent to establish trust, becoming more concise as reliability is demonstrated, and re-expanding situationally when tasks are ambiguous, novel, or high-stakes.
These findings inform design considerations for adaptive feedback with potential transferability beyond driving and in-car assistants, to other primary-task interactions and dual-task contexts where cognitive channels permit intermediate communication.
\newpage

\bibliographystyle{ACM-Reference-Format}
\bibliography{base}

@String{Computing = "Computing" }

@String{Computer = "{IEEE} Computer" }

@String{Springer = "Springer-Verlag" }

@article{wickens2008multiple,
  title={Multiple resources and mental workload},
  author={Wickens, Christopher D.},
  journal={Human Factors},
  volume={50},
  number={3},
  pages={449--455},
  year={2008},
  publisher={SAGE Publications},
doi={10.1518/001872008X288394}
}

@article{horrey2006examining,
  title={Examining the impact of cell phone conversations on driving using meta-analytic techniques},
  author={Horrey, William J and Wickens, Christopher D},
  journal={Human Factors},
  volume={48},
  number={1},
  pages={196--205},
  year={2006},
  publisher={SAGE Publications},
doi={10.1518/001872006776412135}
}

@inproceedings{brewster1994guidelines,
  title={Guidelines for the creation of earcons},
  author={Brewster, Stephen A. and Wright, Peter C. and Edwards, Alistair D.N.},
  booktitle={Proceedings of HCI'94},
  pages={747--759},
  year={1994},
  publisher={Cambridge University Press}
}

@article{burnett2013visual,
  title={On-the-move and in your car: An overview of HCI issues for in-car computing},
  author={Burnett, Gary and Crundall, Elizabeth and Large, David and Lawson, Glyn and de Harder, Liesje},
  journal={International Journal of Mobile Human Computer Interaction},
  volume={5},
  number={1},
  pages={1--21},
  year={2013},
  publisher={IGI Global},
doi={10.4018/jmhci.2009010104}
}

@article{oviatt2000multimodal,
  title={Multimodal interfaces that process what comes naturally},
  author={Oviatt, Sharon and Cohen, Phil},
  journal={Communications of the ACM},
  volume={43},
  number={3},
  pages={45--53},
  year={2000},
  publisher={ACM},
doi={10.1145/330534.330538}
}

@article{shneidermanResponseTimeDisplay1984,
  title   = {Response Time and Display Rate in Human Performance with Computers},
  author  = {Shneiderman, Ben},
  year    = {1984},
  journal = {ACM Computing Surveys},
  volume  = {16},
  number  = {3},
  pages   = {265--285},
  doi     = {10.1145/2514.2517},
}

@inproceedings{funkUsableAcceptableResponse2020,
author = {Funk, Markus and Cunningham, Carie and Kanver, Duygu and Saikalis, Christopher and Pansare, Rohan},
title = {Usable and Acceptable Response Delays of Conversational Agents in Automotive User Interfaces},
year = {2020},
isbn = {9781450380652},
publisher = {Association for Computing Machinery},
address = {New York, USA},
url = {https://doi.org/10.1145/3409120.3410651},
doi = {10.1145/3409120.3410651},
booktitle = {12th International Conference on Automotive User Interfaces and Interactive Vehicular Applications},
pages = {262–269},
numpages = {8},
location = {Virtual Event, DC, USA},
series = {AutomotiveUI '20}
}

@article{strayer2015assessing_cognitive_distraction_in_the_automobile,
  title    = {Assessing Cognitive Distraction in the Automobile},
  volume   = {57},
  issn     = {1547-8181},
  doi      = {10.1177/0018720815575149},
  language = {eng},
  number   = {8},
  journal  = {Human Factors},
  author   = {Strayer, David L. and Turrill, Jonna and Cooper, Joel M. and Coleman, James R. and Medeiros-Ward, Nathan and Biondi, Francesco},
  year     = {2015},
  pmid     = {26534847},
  pages    = {1300--1324}
}

@article{strayerAssessingVisualCognitive2019,
  title    = {Assessing the Visual and Cognitive Demands of In-Vehicle Information Systems},
  author   = {David L. Strayer and Joel M. Cooper and Rachel M. Goethe and Madeleine M. McCarty and Douglas J. Getty and Francesco Biondi},
  year     = {2019},
  journal  = {Cognitive Research: Principles and Implications},
  volume   = {4},
  number   = {1},
  pages    = {18},
  issn     = {2365-7464},
  doi      = {10.1186/s41235-019-0166-3},
}

@inproceedings{myersImportancePercentdoneProgress1985,
  title     = {The Importance of Percent-Done Progress Indicators for Computer-Human Interfaces},
  booktitle = {Proceedings of the {{SIGCHI Conference}} on {{Human Factors}} in {{Computing Systems}}},
  author    = {Myers, Brad A.},
  year      = {1985},
  month     = apr,
  series    = {{{CHI}} '85},
  pages     = {11--17},
  publisher = {Association for Computing Machinery},
  address   = {New York, USA},
  doi       = {10.1145/317456.317459},
}

@inproceedings{hart2006nasa-tlx-20-years-later,
author = {Sandra G. Hart},
title ={Nasa-Task Load Index (NASA-TLX); 20 Years Later},
booktitle = {Proceedings of the Human Factors and Ergonomics Society Annual Meeting},
volume = {50},
pages = {904-908},
year = {2006},
doi = {10.1177/154193120605000909},
publisher = {Sage publications},
address = {Los Angeles, USA},
}

@book{nielsen1994usability,
author = {Nielsen, Jakob},
title = {Usability Engineering},
year = {1994},
isbn = {9780080520292},
publisher = {Morgan Kaufmann Publishers},
address = {San Francisco, USA}
}

@inproceedings{porcheron2018voiceinterfaces,
author = {Porcheron, Martin and Fischer, Joel E. and Reeves, Stuart and Sharples, Sarah},
title = {Voice Interfaces in Everyday Life},
year = {2018},
isbn = {9781450356206},
publisher = {Association for Computing Machinery},
address = {New York, USA},
url = {https://doi.org/10.1145/3173574.3174214},
doi = {10.1145/3173574.3174214},
booktitle = {Proceedings of the 2018 CHI Conference on Human Factors in Computing Systems},
pages = {1–12},
numpages = {12},
location = {Montreal QC, Canada},
series = {CHI '18}
}

@techreport{ADFTW2006,
  author       = {{Group ADFTW}},
  title        = {Statement of principles, criteria and verification procedures on driver interactions with advanced in-vehicle information and communication systems},
  institution  = {Alliance of Automobile Manufacturers},
  year         = {2006},
  type         = {Technical Report}
}

@article{BEANLAND201399,
title = {Driver inattention and driver distraction in serious casualty crashes: Data from the Australian National Crash In-depth Study},
journal = {Accident Analysis \& Prevention},
volume = {54},
pages = {99-107},
year = {2013},
issn = {0001-4575},
doi = {https://doi.org/10.1016/j.aap.2012.12.043},
url = {https://www.sciencedirect.com/science/article/pii/S000145751300047X},
author = {Vanessa Beanland and Michael Fitzharris and Kristie L. Young and Michael G. Lenné}
}

@article{white_risk_2004,
	title = {Risk {Perceptions} of {Mobile} {Phone} {Use} {While} {Driving}},
	volume = {24},
	copyright = {http://onlinelibrary.wiley.com/termsAndConditions\#vor},
	issn = {0272-4332, 1539-6924},
	url = {https://onlinelibrary.wiley.com/doi/10.1111/j.0272-4332.2004.00434.x},
	doi = {10.1111/j.0272-4332.2004.00434.x},
	language = {en},
	number = {2},
	urldate = {2025-09-03},
	journal = {Risk Analysis},
	author = {White, Mathew P. and Eiser, J. Richard and Harris, Peter R.},
	year = {2004},
	pages = {323--334},
}

@article{braun_visualizing_2019,
	title = {Visualizing natural language interaction for conversational in-vehicle information systems to minimize driver distraction},
	volume = {13},
	issn = {1783-7677, 1783-8738},
	url = {http://link.springer.com/10.1007/s12193-019-00301-2},
	doi = {10.1007/s12193-019-00301-2},
	language = {en},
	number = {2},
	urldate = {2025-09-03},
	journal = {Journal on Multimodal User Interfaces},
	author = {Braun, Michael and Broy, Nora and Pfleging, Bastian and Alt, Florian},
	year = {2019},
	pages = {71--88},
}

@inproceedings{maslych2025mitigatingresponsedelays,
author = {Maslych, Mykola and Katebi, Mohammadreza and Lee, Christopher and Hmaiti, Yahya and Ghasemaghaei, Amirpouya and Pumarada, Christian and Palmer, Janneese and Segarra Martinez, Esteban and Emporio, Marco and Snipes, Warren and McMahan, Ryan P. and LaViola Jr., Joseph J.},
title = {Mitigating Response Delays in Free-Form Conversations with LLM-powered Intelligent Virtual Agents},
year = {2025},
isbn = {9798400715273},
publisher = {Association for Computing Machinery},
address = {New York, USA},
doi = {10.1145/3719160.3736636},
booktitle = {Proceedings of the 7th ACM Conference on Conversational User Interfaces},
number = {49},
pages = {15},
location = {},
series = {CUI '25}
}

@article{strayer_talking_2016,
	title = {Talking to your car can drive you to distraction},
	volume = {1},
	issn = {2365-7464},
	url = {http://cognitiveresearchjournal.springeropen.com/articles/10.1186/s41235-016-0018-3},
	doi = {10.1186/s41235-016-0018-3},
	language = {en},
	number = {1},
	urldate = {2025-09-03},
	journal = {Cognitive Research: Principles and Implications},
	author = {Strayer, David L. and Cooper, Joel M. and Turrill, Jonna and Coleman, James R. and Hopman, Rachel J.},
	year = {2016},
	pages = {16}
}

@inproceedings{zhang2024explaining,
author = {Zhang, Zhengquan and Tsiakas, Konstantinos and Schneegass, Christina},
title = {Explaining the Wait: How Justifying Chatbot Response Delays Impact User Trust},
year = {2024},
isbn = {9798400705113},
publisher = {Association for Computing Machinery},
address = {New York, USA},
url = {https://doi.org/10.1145/3640794.3665550},
doi = {10.1145/3640794.3665550},
booktitle = {Proceedings of the 6th ACM Conference on Conversational User Interfaces},
articleno = {27},
numpages = {16},
location = {Luxembourg, Luxembourg},
series = {CUI '24}
}

@inproceedings{cho2025persistent,
author = {Cho, Hyunsung and Fashimpaur, Jacqui and Sendhilnathan, Naveen and Browder, Jonathan and Lindlbauer, David and Jonker, Tanya R. and Todi, Kashyap},
title = {Persistent Assistant: Seamless Everyday AI Interactions via Intent Grounding and Multimodal Feedback},
year = {2025},
isbn = {9798400713941},
publisher = {Association for Computing Machinery},
address = {New York, NY, USA},
url = {https://doi.org/10.1145/3706598.3714317},
doi = {10.1145/3706598.3714317},
booktitle = {Proceedings of the 2025 CHI Conference on Human Factors in Computing Systems},
articleno = {59},
numpages = {19},
location = {
},
series = {CHI '25}
}

@article{atchley2004conversationlimitsfieldofview,
author = {Paul Atchley and Jeff Dressel},
title ={Conversation Limits the Functional Field of View},
journal = {Human Factors},
volume = {46},
number = {4},
pages = {664-673},
year = {2004},
doi = {10.1518/hfes.46.4.664.56808},
 publisher={SAGE Publications Sage UK: London, England}
}

@article{hoff2015trust,
author = {Kevin A. Hoff and Masooda Bashir},
title ={Trust in Automation: Integrating Empirical Evidence on Factors That Influence Trust},
journal = {Human Factors},
volume = {57},
number = {3},
pages = {407-434},
year = {2015},
doi = {10.1177/0018720814547570}, 
}

@article{lee2004trust,
author = {John D. Lee and Katrina A. See},
title ={Trust in Automation: Designing for Appropriate Reliance},
journal = {Human Factors},
volume = {46},
number = {1},
pages = {50-80},
year = {2004},
doi = {10.1518/hfes.46.1.50\_30392},
}

@inproceedings{millerResponseTimeMancomputer1968,
  title     = {Response Time in Man-Computer Conversational Transactions},
  booktitle = {Proceedings of the {{December}} 9-11, 1968, Fall Joint Computer Conference, Part {{I}}},
  author    = {Miller, Robert B.},
  year      = {1968},
  series    = {{{AFIPS}} '68 ({{Fall}}, Part {{I}})},
  pages     = {267--277},
  publisher = {Association for Computing Machinery},
  address   = {New York, USA},
  doi       = {10.1145/1476589.1476628},
  urldate   = {2025-07-17}
}

@inproceedings{schick2023toolformer,
author = {Schick, Timo and Dwivedi-Yu, Jane and Dess\'{\i}, Roberto and Raileanu, Roberta and Lomeli, Maria and Hambro, Eric and Zettlemoyer, Luke and Cancedda, Nicola and Scialom, Thomas},
title = {Toolformer: language models can teach themselves to use tools},
year = {2023},
publisher = {Curran Associates Inc.},
address = {Red Hook, USA},
booktitle = {Proceedings of the 37th International Conference on Neural Information Processing Systems},
articleno = {2997},
numpages = {13},
location = {New Orleans, USA},
series = {NIPS '23}
}

@article{wang_survey_2024,
	title = {A survey on large language model based autonomous agents},
	volume = {18},
	issn = {2095-2228, 2095-2236},
	url = {https://link.springer.com/10.1007/s11704-024-40231-1},
	doi = {10.1007/s11704-024-40231-1},
	language = {en},
	number = {6},
	urldate = {2025-09-03},
	journal = {Frontiers of Computer Science},
	author = {Wang, Lei and Ma, Chen and Feng, Xueyang and Zhang, Zeyu and Yang, Hao and Zhang, Jingsen and Chen, Zhiyuan and Tang, Jiakai and Chen, Xu and Lin, Yankai and Zhao, Wayne X. and Wei, Zhewei and Wen, Jirong},
	year = {2024},
	pages = {186345},
}

@book{norman2002design,
  address     = {New York, USA},
  author      = {Norman, Donald A.},
  description = {The Design of Everyday Things: Amazon.de: Don Norman: Englische Bücher},
  publisher   = {Basic Books},
  refid       = {215302602},
  title       = {The design of everyday things},
  year        = {2002}
}

@book{johnson1986mental,
author = {Philip N. Johnson-Laird},
title = {Mental models: towards a cognitive science of language, inference, and consciousness},
year = {1986},
publisher = {Harvard University Press},
address = {USA}
}

@article{steyversWhatLargeLanguage2025,
  title         = {What {{Large Language Models Know}} and {{What People Think They Know}}},
  author        = {Steyvers, Mark and Tejeda, Heliodoro and Kumar, Aakriti and Belem, Catarina and Karny, Sheer and Hu, Xinyue and Mayer, Lukas and Smyth, Padhraic},
  year          = {2025},
  journal       = {Nature Machine Intelligence},
  volume        = {7},
  number        = {2},
  pages         = {221--231},
  doi           = {10.1038/s42256-024-00976-7},
}

@misc{bansal2024challengeshumanagentcommunication,
      title={Challenges in Human-Agent Communication}, 
      author={Gagan Bansal and Jennifer W. Vaughan and Saleema Amershi and Eric Horvitz and Adam Fourney and Hussein Mozannar and Victor Dibia and Daniel S. Weld},
      year={2024},
doi={10.48550/arXiv.2412.10380},
}

@inproceedings{yankelovich1995ambigoussilence,
author = {Yankelovich, Nicole and Levow, Gina-Anne and Marx, Matt},
title = {Designing SpeechActs: issues in speech user interfaces},
year = {1995},
isbn = {0201847051},
publisher = {ACM Press/Addison-Wesley Publishing Co.},
address = {USA},
url = {https://doi.org/10.1145/223904.223952},
doi = {10.1145/223904.223952},
booktitle = {Proceedings of the SIGCHI Conference on Human Factors in Computing Systems},
pages = {369–376},
numpages = {8},
location = {Denver, USA},
series = {CHI '95}
}

@ARTICLE{axelsson2022modelingfeedback,
  
AUTHOR={Axelsson, Agnes  and Buschmeier, Hendrik  and Skantze, Gabriel },
         
TITLE={Modeling Feedback in Interaction With Conversational Agents—A Review},
        
JOURNAL={Frontiers in Computer Science},
        
VOLUME={4},

YEAR={2022},

PAGES={744574},

URL={https://www.frontiersin.org/journals/computer-science/articles/10.3389/fcomp.2022.744574},

DOI={10.3389/fcomp.2022.744574},

ISSN={2624-9898}
}

@incollection{clark191grounding,
	author = {Herbert H. Clark and Susan E. Brennan},
	booktitle = {Perspectives on Socially Shared Cognition},
	editor = {Lauren Resnick and Levine B. and M. John and Stephanie Teasley and D.},
	pages = {13--1991},
	publisher = {American Psychological Association},
	title = {Grounding in Communication},
	year = {1991},
doi = {10.1037/10096-006},
}

@incollection{brennan_grounding_1998,
  author       = {Brennan, Susan E.},
  title        = {The Grounding Problem in Conversations With and Through Computers},
  booktitle    = {Social and Cognitive Approaches to Interpersonal Communication},
  editor       = {Fussell, Susan R. and Kreuz, Roger J.},
  publisher    = {Lawrence Erlbaum},
  address      = {Hillsdale, NJ},
  year         = {1998},
  edition      = {1st},
  pages        = {201--225},
  isbn         = {978-1-315-807-817},
}

@article{SREEDHARAN2021103558,
title = {Foundations of explanations as model reconciliation},
journal = {Artificial Intelligence},
volume = {301},
pages = {103558},
year = {2021},
issn = {0004-3702},
doi = {https://doi.org/10.1016/j.artint.2021.103558},
url = {https://www.sciencedirect.com/science/article/pii/S0004370221001090},
author = {Sarath Sreedharan and Tathagata Chakraborti and Subbarao Kambhampati}
}

@article{HOSSEINI2025100399,
title = {The role of agentic AI in shaping a smart future: A systematic review},
journal = {Array},
volume = {26},
pages = {100399},
year = {2025},
issn = {2590-0056},
doi = {https://doi.org/10.1016/j.array.2025.100399},
url = {https://www.sciencedirect.com/science/article/pii/S2590005625000268},
author = {Soodeh Hosseini and Hossein Seilani}
}

@article{allwood1992semantics,
  title={On the semantics and pragmatics of linguistic feedback},
  author={Allwood, Jens and Nivre, Joakim and Ahls{\'e}n, Elisabeth},
  journal={Journal of semantics},
  volume={9},
  number={1},
  pages={1--26},
  year={1992},
doi="10.1093/jos/9.1.1",
  publisher={Oxford University Press}
}

@book{clark1996using,
  title={Using language},
  author={Clark, Herbert H.},
  year={1996},
  publisher={Cambridge university press},
doi={10.1017/S0022226798217361},
   address = {Cambridge},
pages = {167--222},
volume = {35},
}

@inproceedings{kirmayr2025carmem,
  title={CarMem: Enhancing Long-Term Memory in LLM Voice Assistants through Category-Bounding},
  author={Kirmayr, Johannes and Stappen, Lukas and Schneider, Phillip and Matthes, Florian and Andre, Elisabeth},
  booktitle={Proceedings of the 31st International Conference on Computational Linguistics: Industry Track},
  pages={343--357},
  year={2025},
publisher= {Association for Computational Linguistics},
address = {Abu Dhabi, UAE},
}

@INPROCEEDINGS{stappen2023genaiautomotive,
  author={Stappen, Lukas and Dillmann, Jeremy and Striegel, Serena and Vögel, Hans-Jörg and Flores-Herr, Nicolas and Schuller, Björn W.},
  booktitle={2023 IEEE 26th International Conference on Intelligent Transportation Systems (ITSC)}, 
address={Bilbao, Spain},
  title={Integrating Generative Artificial Intelligence in Intelligent Vehicle Systems}, 
  year={2023},
  volume={},
  number={},
  pages={5790-5797},
  doi={10.1109/ITSC57777.2023.10422003}}

@inproceedings{jaber2024cookingwithagents,
author = {Jaber, Razan and Zhong, Sabrina and Kuoppam\"{a}ki, Sanna and Hosseini, Aida and Gessinger, Iona and Brumby, Duncan P. and Cowan, Benjamin R. and Mcmillan, Donald},
title = {Cooking With Agents: Designing Context-aware Voice Interaction},
year = {2024},
publisher = {Association for Computing Machinery},
address = {New York, USA},
doi = {10.1145/3613904.3642183},
booktitle = {Proceedings of the 2024 CHI Conference on Human Factors in Computing Systems},
articleno = {551},
numpages = {13},
location = {Honolulu, USA},
series = {CHI '24}
}

@ARTICLE{acharya2025agenticai,
  author={Acharya, Deepak B. and Kuppan, Karthigeyan and Divya, B.},
  journal={IEEE Access}, 
  title={Agentic AI: Autonomous Intelligence for Complex Goals—A Comprehensive Survey}, 
  year={2025},
  volume={13},
  number={},
  pages={18912-18936},
  doi={10.1109/ACCESS.2025.3532853}}

@misc{openai_deep_research,
	title = {Introducing deep research},
	url = {https://openai.com/index/introducing-deep-research/},
        lastaccessed = "September 5, 2025",
	author = {OpenAI},
	month = feb,
	year = {2025},
}

@Article{airportchatbot,
AUTHOR = {Auer, Isabel and Schlögl, Stephan and Glowka, Gundula},
TITLE = {Chatbots in Airport Customer Service—Exploring Use Cases and Technology Acceptance},
JOURNAL = {Future Internet},
VOLUME = {16},
YEAR = {2024},
NUMBER = {5},
PAGES = {175},
URL = {https://www.mdpi.com/1999-5903/16/5/175},
ISSN = {1999-5903},
DOI = {10.3390/fi16050175}
}

@misc{chen2024travelagentaiassistantpersonalized,
      title={TravelAgent: An AI Assistant for Personalized Travel Planning}, 
      author={Aili Chen and Xuyang Ge and Ziquan Fu and Yanghua Xiao and Jiangjie Chen},
      year={2024},
      doi={10.48550/arXiv.2409.08069}, 
}

@misc{openai2024gpt4ocard,
      title={GPT-4o System Card}, 
      author={OpenAI},
      year={2024},
      doi={10.48550/arXiv.2410.21276}, 
}

@misc{sonnet_system_card,
	title = {System Card: Claude Opus 4 \& Claude Sonnet 4},
	url = {https://www-cdn.anthropic.com/6d8a8055020700718b0c49369f60816ba2a7c285.pdf},
        lastaccessed = "September 5, 2025",
	author = {Anthropic},
	month = may,
	year = {2025},
}

@misc{comanici2025gemini25pushingfrontier,
      title={Gemini 2.5: Pushing the Frontier with Advanced Reasoning, Multimodality, Long Context, and Next Generation Agentic Capabilities}, 
      author={Google Gemini Team},
      year={2025},
      doi={10.48550/arXiv.2507.06261}, 
}

@article{mcgrathMeasuringTrustArtificial2025,
  title      = {Measuring Trust in Artificial Intelligence: Validation of an Established Scale and Its Short Form},
  shorttitle = {Measuring Trust in Artificial Intelligence},
  author     = {McGrath, Melanie J. and Lack, Oliver and Tisch, James and Duenser, Andreas},
  year       = {2025},
  journal    = {Frontiers in Artificial Intelligence},
  volume     = {8},
  publisher  = {Frontiers},
  doi        = {10.3389/frai.2025.1582880}
}

@incollection{HART1988139,
title = {Development of NASA-TLX (Task Load Index): Results of Empirical and Theoretical Research},
editor = {Peter A. Hancock and Najmedin Meshkati},
series = {Advances in Psychology},
publisher = {North-Holland},
volume = {52},
pages = {139-183},
year = {1988},
booktitle = {Human Mental Workload},
issn = {0166-4115},
doi = {https://doi.org/10.1016/S0166-4115(08)62386-9},
author = {Sandra G. Hart and Lowell E. Staveland},
}

@article{ueqplus,
  author = {Martin Schrepp and J\"org Thomaschewski},
  title = {Design and Validation of a Framework for the Creation of User Experience Questionnaires},
  year = {2019},
  journal = {International Journal of Interactive Multimedia and Artificial Intelligence},
  volume = {5},
  number = {7},
  pages = {88-95},
doi= {10.9781/ijimai.2019.06.006},
}

@ARTICLE{lukasik2025fromrobotstochatbots,
  
AUTHOR={Łukasik, Albert  and Gut, Arkadiusz },
         
TITLE={From robots to chatbots: unveiling the dynamics of human-AI interaction},
        
JOURNAL={Frontiers in Psychology},
        
VOLUME={Volume 16 - 2025},

YEAR={2025},

URL={https://www.frontiersin.org/journals/psychology/articles/10.3389/fpsyg.2025.1569277},

DOI={10.3389/fpsyg.2025.1569277},

ISSN={1664-1078}}

@inproceedings{kraus2022includingsocial,
author = {Kraus, Matthias and Wagner, Nicolas and Untereiner, Nico and Minker, Wolfgang},
title = {Including Social Expectations for Trustworthy Proactive Human-Robot Dialogue},
year = {2022},
isbn = {9781450392075},
publisher = {Association for Computing Machinery},
address = {New York, USA},
url = {https://doi.org/10.1145/3503252.3531294},
doi = {10.1145/3503252.3531294},
booktitle = {Proceedings of the 30th ACM Conference on User Modeling, Adaptation and Personalization},
pages = {23–33},
numpages = {11},
location = {Barcelona, Spain},
series = {UMAP '22}
}

@misc{tool_preamples,
	title = {GPT-5 prompting guide - Tool preambles},
	url = {https://cookbook.openai.com/examples/gpt-5/gpt-5_prompting_guide#tool-preambles},
        lastaccessed = "September 8, 2025",
	author = {OpenAI},
	month = aug,
	year = {2025},
}

@inproceedings{ueno2022trustinhumanAIinteraction,
author = {Ueno, Takane and Sawa, Yuto and Kim, Yeongdae and Urakami, Jacqueline and Oura, Hiroki and Seaborn, Katie},
title = {Trust in Human-AI Interaction: Scoping Out Models, Measures, and Methods},
year = {2022},
isbn = {9781450391566},
publisher = {Association for Computing Machinery},
address = {New York, USA},
doi = {10.1145/3491101.3519772},
booktitle = {Extended Abstracts of the 2022 CHI Conference on Human Factors in Computing Systems},
articleno = {254},
numpages = {7},
location = {New Orleans, USA},
series = {CHI EA '22}
}

@misc{packer2024memgptllmsoperatingsystems,
      title={MemGPT: Towards LLMs as Operating Systems}, 
      author={Charles Packer and Sarah Wooders and Kevin Lin and Vivian Fang and Shishir G. Patil and Ion Stoica and Joseph E. Gonzalez},
      year={2024},
      doi={10.48550/arXiv.2310.08560}, 
}

@misc{kalai2025languagemodelshallucinate,
      title={Why Language Models Hallucinate}, 
      author={Adam T. Kalai and Ofir Nachum and Santosh S. Vempala and Edwin Zhang},
      year={2025},
      doi={10.48550/arXiv.2509.04664}, 
}

@inproceedings{zhu-etal-2023-calibration,
    title = "On the Calibration of Large Language Models and Alignment",
    author = "Zhu, Chiwei  and
      Xu, Benfeng  and
      Wang, Quan  and
      Zhang, Yongdong  and
      Mao, Zhendong",
    editor = "Bouamor, Houda  and
      Pino, Juan  and
      Bali, Kalika",
    booktitle = "Findings of the Association for Computational Linguistics: EMNLP 2023",
    month = dec,
    year = "2023",
    address = "Singapore",
    publisher = "Association for Computational Linguistics",
    url = "https://aclanthology.org/2023.findings-emnlp.654/",
    doi = "10.18653/v1/2023.findings-emnlp.654",
    pages = "9778--9795"
}

@article{song2022role,
  title={The role of the human-robot interaction in consumers’ acceptance of humanoid retail service robots},
  author={Song, Christina S. and Kim, Youn-Kyung},
  journal={Journal of Business Research},
  volume={146},
  number={2},
  pages={489--503},
  year={2022},
  publisher={Elsevier},
doi= {10.1016/j.jbusres.2022.03.087}
}

@misc{cursor,
	title = {The AI Code Editor},
	url = {https://cursor.com/home},
	urldate = {2025-09-10},
        lastaccessed = "September 10, 2025",
	author = {Cursor},
	month = sep,
	year = {2025},
}

@misc{perplexity,
	title = {Perplexity},
	url = {https://www.perplexity.ai/},
	urldate = {2025-09-10},
        lastaccessed = "September 10, 2025",
	author = {Perplexity AI},
	year = {2025},
}

@misc{manus,
	title = {Manus},
	url = {https://manus.im/?index=1},
	urldate = {2025-09-10},
        lastaccessed = "September 10, 2025",
	author = {Manus AI},
	month = sep,
	year = {2025},
}

@inproceedings{xu2015optimo,
  title={Optimo: Online probabilistic trust inference model for asymmetric human-robot collaborations},
  author={Xu, Anqi and Dudek, Gregory},
  booktitle={Proceedings of the tenth annual ACM/IEEE international conference on human-robot interaction},
  pages={221--228},
  year={2015}
}

@article{mehrotra2024trust,
author = {Mehrotra, Siddharth and Degachi, Chadha and Vereschak, Oleksandra and Jonker, Catholijn M. and Tielman, Myrthe L.},
title = {A Systematic Review on Fostering Appropriate Trust in Human-AI Interaction: Trends, Opportunities and Challenges},
year = {2024},
publisher = {Association for Computing Machinery},
address = {New York, USA},
volume = {1},
number = {4},
doi = {10.1145/3696449},
journal = {ACM Journal on Responsible Computing},
articleno = {26},
numpages = {45}
}

@book{blandford2016qualitative,
	series = {Synthesis {Lectures} on {Human}-{Centered} {Informatics}},
	title = {Qualitative {HCI} {Research}: {Going} {Behind} the {Scenes}},
	copyright = {https://www.springer.com/tdm},
	isbn = {978-3-031-01089-7 978-3-031-02217-3},
	shorttitle = {Qualitative {HCI} {Research}},
	url = {https://link.springer.com/10.1007/978-3-031-02217-3},
	language = {en},
	urldate = {2025-09-11},
	publisher = {Springer, Cham},
	author = {Blandford, Ann and Furniss, Dominic and Makri, Stephann},
	year = {2016},
	doi = {10.1007/978-3-031-02217-3},
}

@misc{trustPairGoogle,
	title = {Trust + Explanations Help people recover from errors},
	url = {https://pair.withgoogle.com/guidebook/chapters/trust-and-explanations/understanding-trust-in-your-product},
	urldate = {2025-09-10},
        lastaccessed = "September 10, 2025",
	author = {Google},
	month = sep,
	year = {2025},
}

@article{liu2021inaiwetrust,
    author = {Liu, Bingjie},
    title = {In AI We Trust? Effects of Agency Locus and Transparency on Uncertainty Reduction in Human–AI Interaction},
    journal = {Journal of Computer-Mediated Communication},
    volume = {26},
    number = {6},
    pages = {384-402},
    year = {2021},
    month = {09},
    issn = {1083-6101},
    doi = {10.1093/jcmc/zmab013},
    url = {https://doi.org/10.1093/jcmc/zmab013},
    eprint = {https://academic.oup.com/jcmc/article-pdf/26/6/384/41139653/zmab013.pdf},
}

@inproceedings{wei2022cot,
author = {Wei, Jason and Wang, Xuezhi and Schuurmans, Dale and Bosma, Maarten and Ichter, Brian and Xia, Fei and Chi, Ed H. and Le, Quoc V. and Zhou, Denny},
title = {Chain-of-thought prompting elicits reasoning in large language models},
year = {2022},
isbn = {9781713871088},
publisher = {Curran Associates Inc.},
address = {Red Hook, NY, USA},
booktitle = {Proceedings of the 36th International Conference on Neural Information Processing Systems},
articleno = {1800},
numpages = {14},
location = {New Orleans, LA, USA},
series = {NIPS '22}
}

@article{vossing_designing_2022,
	title = {Designing {Transparency} for {Effective} {Human}-{AI} {Collaboration}},
	volume = {24},
	issn = {1387-3326, 1572-9419},
	url = {https://link.springer.com/10.1007/s10796-022-10284-3},
	doi = {10.1007/s10796-022-10284-3},
	language = {en},
	number = {3},
	urldate = {2025-09-12},
	journal = {Information Systems Frontiers},
	author = {Vössing, Michael and Kühl, Niklas and Lind, Matteo and Satzger, Gerhard},
	month = jun,
	year = {2022},
	pages = {877--895},
}

@inproceedings{xie2019robot,
author = {Xie, Yaqi and Bodala, Indu P and Ong, Desmond C. and Hsu, David and Soh, Harold},
title = {Robot capability and intention in trust-based decisions across tasks},
year = {2020},
isbn = {9781538685556},
publisher = {IEEE Press},
booktitle = {Proceedings of the 14th ACM/IEEE International Conference on Human-Robot Interaction},
pages = {39–47},
numpages = {9},
location = {Daegu, Republic of Korea},
series = {HRI '19}
}

@article{vereshak2021howtoevaluatetrust,
author = {Vereschak, Oleksandra and Bailly, Gilles and Caramiaux, Baptiste},
title = {How to Evaluate Trust in AI-Assisted Decision Making? A Survey of Empirical Methodologies},
year = {2021},
issue_date = {October 2021},
publisher = {Association for Computing Machinery},
address = {New York, NY, USA},
volume = {5},
number = {CSCW2},
url = {https://doi.org/10.1145/3476068},
doi = {10.1145/3476068},
journal = {Proc. ACM Hum.-Comput. Interact.},
month = oct,
articleno = {327},
numpages = {39}
}

@article{VIRVOU2024120759,
title = {VIRTSI: A novel trust dynamics model enhancing Artificial Intelligence collaboration with human users – Insights from a ChatGPT evaluation study},
journal = {Information Sciences},
volume = {675},
pages = {120759},
year = {2024},
issn = {0020-0255},
doi = {https://doi.org/10.1016/j.ins.2024.120759},
url = {https://www.sciencedirect.com/science/article/pii/S002002552400673X},
author = {Maria Virvou and George A. Tsihrintzis and Evangelia-Aikaterini Tsichrintzi}
}

@inproceedings{kobalczyk2025activetaskdisambiguationllms,
      title={Active Task Disambiguation with LLMs}, 
      author={Katarzyna Kobalczyk and Nicolas Astorga and Tennison Liu and Mihaela van der Schaar},
      year={2025},
      doi={10.48550/arXiv.2502.04485}, 
booktitle={The Thirteenth International Conference on Learning Representations},
address={Singapore}
}

@inproceedings{almada2019humanintervention,
author = {Almada, Marco},
title = {Human intervention in automated decision-making: Toward the construction of contestable systems},
year = {2019},
isbn = {9781450367547},
publisher = {Association for Computing Machinery},
address = {New York, NY, USA},
url = {https://doi.org/10.1145/3322640.3326699},
doi = {10.1145/3322640.3326699},
booktitle = {Proceedings of the Seventeenth International Conference on Artificial Intelligence and Law},
pages = {2–11},
numpages = {10},
location = {Montreal, QC, Canada},
series = {ICAIL '19}
}

@inproceedings{sterz2024onthequest,
author = {Sterz, Sarah and Baum, Kevin and Biewer, Sebastian and Hermanns, Holger and Lauber-R\"{o}nsberg, Anne and Meinel, Philip and Langer, Markus},
title = {On the Quest for Effectiveness in Human Oversight: Interdisciplinary Perspectives},
year = {2024},
isbn = {9798400704505},
publisher = {Association for Computing Machinery},
address = {New York, NY, USA},
url = {https://doi.org/10.1145/3630106.3659051},
doi = {10.1145/3630106.3659051},
booktitle = {Proceedings of the 2024 ACM Conference on Fairness, Accountability, and Transparency},
pages = {2495–2507},
numpages = {13},
location = {Rio de Janeiro, Brazil},
series = {FAccT '24}
}

@inproceedings{he2025planthenexecute,
author = {He, Gaole and Demartini, Gianluca and Gadiraju, Ujwal},
title = {Plan-Then-Execute: An Empirical Study of User Trust and Team Performance When Using LLM Agents As A Daily Assistant},
year = {2025},
isbn = {9798400713941},
publisher = {Association for Computing Machinery},
address = {New York, NY, USA},
url = {https://doi.org/10.1145/3706598.3713218},
booktitle = {Proceedings of the 2025 CHI Conference on Human Factors in Computing Systems},
articleno = {414},
numpages = {22}
}

@article{langer_effective_2024,
	title = {Effective Human Oversight of AI-Based Systems: A Signal Detection Perspective on the Detection of Inaccurate and Unfair Outputs},
	volume = {35},
	issn = {1572-8641},
	shorttitle = {Effective {Human} {Oversight} of {AI}-{Based} {Systems}},
	url = {https://link.springer.com/10.1007/s11023-024-09701-0},
	doi = {10.1007/s11023-024-09701-0},
	language = {en},
	number = {1},
	urldate = {2025-12-03},
	journal = {Minds and Machines},
	author = {Langer, Markus and Baum, Kevin and Schlicker, Nadine},
	month = nov,
	year = {2024},
	pages = {1},
}

@article{dietvorst2018overcoming,
  title={Overcoming algorithm aversion: People will use imperfect algorithms if they can (even slightly) modify them},
  author={Dietvorst, Berkeley J and Simmons, Joseph P and Massey, Cade},
  journal={Management science},
  volume={64},
  number={3},
  pages={1155--1170},
  year={2018},
  publisher={Informs}
}

@article{LOEW2023106898,
title = {The impact of speech-based assistants on the driver’s cognitive distraction},
journal = {Accident Analysis \& Prevention},
volume = {179},
pages = {106898},
year = {2023},
issn = {0001-4575},
doi = {https://doi.org/10.1016/j.aap.2022.106898},
author = {Alexandra Loew and Ina Koniakowsky and Yannick Forster and Frederik Naujoks and Andreas Keinath},
}

@inproceedings{sorokin2025collaborating,
author = {Sorokin, Lenja and Huynh, Khanh and Eiband, Malin and Stappen, Lukas and Dillmann, Jeremy},
title = {Collaborating with LLMs Through a Voice and Graphical User Interface},
year = {2025},
isbn = {9798400719707},
publisher = {Association for Computing Machinery},
address = {New York, NY, USA},
doi = {10.1145/3737821.3749555},
booktitle = {Adjunct Proceedings of the 27th International Conference on Mobile Human-Computer Interaction},
articleno = {3},
numpages = {8},
location = {
},
series = {MobileHCI '25 Adjunct}
}

@article{ZHANG2023103958,
title = {Input modality matters: A comparison of touch, speech, and gesture based in-vehicle interaction},
journal = {Applied Ergonomics},
volume = {108},
pages = {103958},
year = {2023},
issn = {0003-6870},
doi = {https://doi.org/10.1016/j.apergo.2022.103958},
author = {Tingru Zhang and Xing Liu and Weisheng Zeng and Da Tao and Guofa Li and Xingda Qu}
}

@article{lo2013developmentofspeechinterfaces,
author = {Lo, Victor Ei-Wen and Green, Paul A.},
title = {Development and Evaluation of Automotive Speech Interfaces: Useful Information from the Human Factors and the Related Literature},
journal = {International Journal of Vehicular Technology},
volume = {2013},
number = {1},
pages = {924170},
doi = {https://doi.org/10.1155/2013/924170},
year = {2013}
}

@misc{kirmayr2026carbenchevaluatingconsistencylimitawareness,
      title={CAR-bench: Evaluating the Consistency and Limit-Awareness of LLM Agents under Real-World Uncertainty}, 
      author={Johannes Kirmayr and Lukas Stappen and Elisabeth André},
      year={2026},
      eprint={2601.22027},
      archivePrefix={arXiv},
      primaryClass={cs.AI},
      url={https://arxiv.org/abs/2601.22027}, 
}

@misc{stappen2026agent2agentthreatssafetycriticalllm,
      title={Agent2Agent Threats in Safety-Critical LLM Assistants: A Human-Centric Taxonomy}, 
      author={Lukas Stappen and Ahmet Erkan Turan and Johann Hagerer and Georg Groh},
      year={2026},
      eprint={2602.05877},
      archivePrefix={arXiv},
      primaryClass={cs.AI},
      url={https://arxiv.org/abs/2602.05877}, 
}

@inproceedings{khanh2025spatial,
author = {Huynh, Khanh and Dillmann, Jeremy and Mayer, Sven},
title = {Spatial Referencing for Large Language Models in Automotive Navigation Tasks},
year = {2025},
isbn = {9798400720154},
publisher = {Association for Computing Machinery},
address = {New York, NY, USA},
url = {https://doi.org/10.1145/3771882.3771917},
doi = {10.1145/3771882.3771917},
booktitle = {Proceedings of the 24th International Conference on Mobile and Ubiquitous Multimedia},
pages = {146–157},
numpages = {12},
location = {
},
series = {MUM '25}
}

% %%
% %% If your work has an appendix, this is the place to put it.
\appendix

\section{Survey}
The following questions are translated from german into english.
\subsection{Demographics \& Technical Familiarity}
\begin{itemize}
    \item Age
    \begin{itemize}
        \item Under 18
        \item 18–24 years
        \item 25–34 years
        \item 35–44 years
        \item 45–54 years
        \item 55–64 years
        \item 65 years and older
    \end{itemize}
    \item How do you describe yourself?
    \begin{itemize}
        \item Male
        \item Female
        \item Non-binary / third gender
        \item Self-description preferred
        \item Prefer not to answer
    \end{itemize}
    \item How familiar are you with the general capabilities and functionalities of Large Language Models (LLMs) such as ChatGPT?
    \begin{itemize}
        \item Not at all familiar
        \item Somewhat familiar
        \item Familiar
        \item Very familiar
        \item Extremely familiar        
    \end{itemize}
    \item How familiar are you with voice assistants such as Alexa, Siri, Google Assistant, etc.?
    \begin{itemize}
        \item Not at all familiar
        \item Somewhat familiar
        \item Familiar
        \item Very familiar
        \item Extremely familiar 
    \end{itemize}
    \item How familiar are you with the companies in-car voice assistant?
    \begin{itemize}
        \item Not at all familiar
        \item Somewhat familiar
        \item Familiar
        \item Very familiar
        \item Extremely familiar 
    \end{itemize}
\end{itemize}

\subsection{Questionnaires}

\subsubsection{Perceived Speed}
\begin{itemize}
    \item How fast or slow did you perceive the system during the task?
    \begin{itemize}
        \item Very slow (1) – Very fast (7)
    \end{itemize}
\end{itemize}

\subsubsection{User Experience - UEQ+ subset}
\begin{itemize}
    \item Attractiveness
    \begin{itemize}
        \item In my opinion, the product is generally:
        \begin{itemize}
            \item annoying (-3) – enjoyable (3)
            \item Bad (-3) – Good (3)
            \item unpleasant (-3) – pleasant (3)
            \item unfriendly (-3) – friendly (3)
        \end{itemize}
        \item The product characteristic described by these terms is for me
        \begin{itemize}
            \item Not important at all (1) - Very important (7)
        \end{itemize}
    \end{itemize}
    \item Dependability
    \begin{itemize}
        \item In my opinion, the reactions of the product to my input and command are:
        \begin{itemize}
            \item unpredictable (-3) – predictable (3)
            \item obstructive (-3) – supportive (3)
            \item not secure (-3) – secure (3)
            \item does not meet expectations (-3) – meets expectations (3)
        \end{itemize}
        \item The product characteristic described by these terms is for me
        \begin{itemize}
            \item Not important at all (1) - Very important (7)
        \end{itemize}
    \end{itemize}
    \item Risk Handling
    \begin{itemize}
        \item I find the application errors and risks which may arise when using the product to be:
        \begin{itemize}
            \item threatening (-3) – harmless (3)
            \item hazardous to health  (-3) – not hazardous to health (3)
            \item damaging (-3) – not damaging (3)
            \item likely to cause collision (-3) – unlikely to cause collision (3)
        \end{itemize}
        \item The product characteristic described by these terms is for me
        \begin{itemize}
            \item Not important at all (1) - Very important (7)
        \end{itemize}
    \end{itemize}
\end{itemize}

\subsubsection{Task Load - NASA-RTLX subset}
Please indicate for each of the dimensions below how demanding the task was for you. Mark on the following scales to what extent you felt challenged or required in the six mentioned dimensions:
\begin{itemize}
    \item Mental Demand
    \begin{itemize}
        \item How much mental and perceptual activity was required (e.g., thinking, deciding, calculating, remembering, observing, searching…)? Was the task easy or demanding, simple or complex, exacting or forgiving?
        \begin{itemize}
            \item Low (0) – High (100)
        \end{itemize}
    \end{itemize}
    \item Temporal Demand
    \begin{itemize}
        \item How much time pressure did you feel due to the rate or pace at which the tasks occured? Was the pace slow and leisurely or rapid and frantic?
        \begin{itemize}
            \item Low (0) – High (100)
        \end{itemize}
    \end{itemize}
    \item Frustration level
    \begin{itemize}
        \item How insecure, discouraged, irritated, stressed and annoyed versus secure, gratified, content, relaxed and complacement did you feel during the task?
        \begin{itemize}
            \item Low (0) – High (100)
        \end{itemize}
    \end{itemize}
\end{itemize}

\subsubsection{User Trust - S-TIAS}
Please indicate the extent to which you agree with the following statements about the voice assistant:
\begin{itemize}
    \item I have confidence in the assistant
    \begin{itemize}
        \item Not at all (1) - Extremely (7)
    \end{itemize}
    \item The system is reliable
    \begin{itemize}
        \item Not at all (1) - Extremely (7)
    \end{itemize}
    \item I can trust the system
    \begin{itemize}
        \item Not at all (1) - Extremely (7)
    \end{itemize}
\end{itemize}

\subsection{Semi-structured Interview Questions}
The following questions were asked while allowing follow-up prompts and clarification questions:
\begin{enumerate}
    \item How much verbal feedback would you like from the system? Consider driving situation, passengers, music, and other distractions.
    \item Should the system notify you when it is uncertain, or decide autonomously? If notified, how should this be communicated?
    \item Which system behaviors or experiences would foster long-term trust?
\end{enumerate}








%TC:endignore
\end{document}